\documentclass[5p,times,number]{elsarticle}
\pdfoutput=1

\usepackage{graphicx,miller,bm,booktabs,float,nicefrac,url,color,microtype,subcaption,lipsum}
\usepackage[export]{adjustbox}
\usepackage{siunitx}[=v2]
\usepackage[version=4]{mhchem}
\usepackage{hyperref}
\usepackage[ruled]{algorithm2e}

\hypersetup{breaklinks,hidelinks}

\DeclareMathAccent{\wtilde}{\mathord}{largesymbols}{"65}
\newcommand*{\matr}[1]{\bm{\mathit{#1}}}

\DeclareSIUnit{\pixel}{pixels}
\DeclareSIUnit{\voxel}{voxels}

\journal{Materials Characterization}

\begin{document}

\begin{frontmatter}

\title{Registration between DCT and EBSD datasets for multiphase microstructures}

\author[a,b]{James A. D. Ball}
\author[c]{Jette Oddershede}
\author[d]{Claire Davis}
\author[d]{Carl Slater}
\author[a]{Mohammed Said}
\author[a]{Himanshu Vashishtha}
\author[b]{Stefan Michalik}
\author[a]{David M. Collins}
\ead{D.M.Collins@bham.ac.uk}
\affiliation[a]{organization={School of Metallurgy and Materials, University of Birmingham},
        addressline={Edgbaston}, 
        city={Birmingham},
        postcode={B15~2TT}, 
        country={United~Kingdom}}
\affiliation[b]{organization={Diamond Light Source Ltd.},
        addressline={Harwell Science and Innovation Campus}, 
        city={Didcot},
        postcode={OX11~0DE}, 
        country={United~Kingdom}}
\affiliation[c]{organization={Xnovo Technology ApS},
    addressline={Galoche Alle 15}, 
    city={Køge},
    postcode={4600}, 
    country={Denmark}}
\affiliation[d]{organization={WMG, University of Warwick}, 
    city={Coventry},
    postcode={CV4~7AL}, 
    country={United~Kingdom}}

\begin{abstract}
The ability to characterise the three-dimensional microstructure of multiphase materials is essential for understanding the interaction between phases and associated materials properties.
Here, laboratory-based diffraction-contrast tomography (lab-based DCT), a recently-established materials characterization technique that can determine grain phases, morphologies, positions and orientations in a voxel-based reconstruction method, was used to map part of a dual-phase steel alloy sample.
To assess the resulting microstructures that were produced by the lab-based DCT technique, an electron backscatter diffraction (EBSD) map was collected within the same sample volume.
To identify the two-dimensional (2D) slice of the three-dimensional (3D) lab-based DCT reconstruction that best corresponded to the 2D EBSD map, a novel registration technique based solely on grain-averaged orientations was developed -- this registration technique requires very little \emph{a priori} knowledge of dataset alignment and can be extended to other techniques that only recover grain-averaged orientation data such as far-field 3D X-ray diffraction microscopy.
Once the corresponding 2D slice was identified in the lab-based DCT dataset, comparisons of phase balance, grain size, shape and texture were performed between lab-based DCT and EBSD techniques.
More complicated aspects of the microstructural morphology such as grain boundary shape and grains less than a critical size were poorly reproduced by the lab-based DCT reconstruction, primarily due to the difference in resolutions of the technique compared with EBSD.
However, lab-based DCT is shown to accurately determine the centre-of-mass position, orientation, and size of the large grains for each phase present, austenite and martensitic ferrite. The results reveals a complex ferrite grain network of similar crystal orientations that are absent from the EBSD dataset. Such detail demonstrates that lab-based DCT, as a technique, shows great promise in the field of multi-phase material characterization.
\end{abstract}


\end{frontmatter}

\section{Introduction}
Understanding the deformation behaviour of multiphase polycrystalline structural alloys, such as $\alpha/\beta$ titanium alloys for compressor discs in aeroengines or high strength dual phase (ferritic–martensitic) steels for automotive, load bearing chassis components, is vital for guiding the design of future materials.
The micromechanical material behaviour is intimately linked to the material microstructure, not only to the phase fractions, but also to the phase specific grain size distributions (i.e. fine-grained/coarse-grained/bimodal, narrow/wide), grain shapes (i.e. equiaxed/needles/plates), and textures.
These features will have an associated distribution throughout the material in 3D; this may be uniform or heterogeneous, often inherited from the prior processing.

Interaction between phases, such as load shedding, is a critical attribute that must be well known for predicting failure initiation and deformation evolution.
For probing the load sharing among phases, experimental techniques such as far-field 3DXRD (3-Dimensional X-ray Diffraction)/HEDM (High Energy X-ray Diffraction Microscopy) have proved to be excellent methods as they are sensitive to the phase, center of mass position, crystallographic orientation, and lattice distortions (and hence grain averaged type-II stress) of every grain, non-destructively, in 3D \cite{sedmak2016,Nair2019}.
Non-destructive experimental techniques to map out the 3D grain structure comprise near-field 3DXRD/HEDM e.g. \cite{Rodek2007,Menasche2020,SHEN2020100852,Li2013512} and synchrotron DCT \cite{ludwig_x-ray_2008,johnson_x-ray_2008,shiozawa_4d_2016,ribart_situ_2023}.
Obtaining such data without the need to access national or international facilities is also possible via lab-based DCT \cite{king_first_2013, mcdonald_non-destructive_2015, keinan_integrated_2018, zhao_correlation_2022}, which is highly attractive if it has the capability to accurately describe microstructures of engineering alloys.

For a multi-phase material, lab-based DCT has been used to good effect to elucidate hydrogen embrittlement in a duplex stainless steel consisting of a dual-phase ferrite-austenite microstructure \citep{eguchi_x-ray_2022}.
Here, an old version of the reconstruction engine, GrainMapper3D™, was used to reconstruct each phase separately.
The software constrained the analysis to provide only grain sizes, center of mass positions, and crystallographic orientations, while the movement of diffraction spots was used, after hydrogen infusion, to qualitatively evaluate stress evolution.

EBSD (electron backscatter diffraction) can be used to characterise a 2D region on the sample surface of a polycrystalline material (e.g. \cite{MIYAMOTO20091120,ZHANG201425,wallis2019,DEAL2021111027,KOKO2023105173}), with comparably better spatial resolution than 3D techniques.
The use of 3D-EBSD, where an individual EBSD scan is acquired with successive serial sectioning (e.g. \cite{GROEBER2006259,Rohrer2010,KALACSKA2020211,DEMOTT2021113394}),  is attractive.
However, being a destructive method, in-situ studies are impossible with 3D-EBSD, which limits its applicability to study deformation.
EBSD is used routinely for investigating multiphase structural materials \cite{EBSD}, while only a very limited number of 3D space filling grain maps of dual phase materials exist, e.g. \cite{Pokharel2020}.

To directly compare grain center-of-mass data to space filling 2D/3D grain maps, or grain maps to each other, a registration approach is needed.
Often, the registration technique used during the comparison is not specified \citep{johnson_x-ray_2008, ribart_situ_2023}.
In some cases, both measurement techniques are performed at the same facility and therefore use the same reference frames and length scales, so a post-mortem registration is unnecessary \citep{ludwig_three-dimensional_2009}.
For cases where algorithmic registration is required, a range of different dataset registration strategies have been employed, such as manual alignment \citep{king_first_2013}, plane fitting using porosity data \citep{syha_validation_2013} and misorientation minimization \citep{quey_direct_2013, renversade_comparison_2016}.
As the reconstructed grain maps are inherently multi-dimensional, and/or multimodal, visualization packages such as \textit{PolyProc} offer dataset filtration and analysis capability \cite{kang_polyproc_2019}.

For microstructures comprising multiple constituent phases, there is no registration algorithm developed to date that can handle a combination of center of mass or space filling data, for each phase present, in both 2D and 3D.
Hence, this study seeks to implement such a registration algorithm and test its performance on a difficult case, namely a two-phase metastable austenitic stainless steel with an austenite and martensitic-ferrite structure mapped by both EBSD and lab-based DCT.
A direct one-to-one comparison between the two methods is considered unreachable due to a morphologically complex, fine-grained dual-phase microstructure of the sample -- some of these microstructural features are outside of the detection limits of the DCT method (\SIrange[range-phrase=--, range-units = single]{10}{40}{\micro\metre} \citep{oddershede_non-destructive_2019}).
However, comparing statistical properties of the 2D-EBSD map to the nearest 2D slice in the 3D-DCT is reachable.
This can be determined by the registration between the datasets.
In this study, the corresponding properties are also derived for a full 3D-DCT volume to ascertain the advantages and disadvantages of EBSD versus DCT for the grain mapping of multiphase samples.

\section{Experimental Method}
\subsection{Material}
The alloy studied was a two-phase austenitic metastable stainless steel alloy with the composition given in Table~\ref{tab:alloy_composition}.
\begin{table}[H]
\centering
\caption{Experimental alloy composition}
\label{tab:alloy_composition}
\begin{tabular}{@{}l|llllllll@{}}
\textbf{Element}     & C    & Ni & Cr & Mn & Si & P    & S    & Fe   \\ \midrule
\textbf{wt.\%} & 0.04 & 7  & 19 & 2  & 1  & 0.04 & 0.03 & Bal.
\end{tabular}
\end{table}
The alloy was cast as a \SI{10}{\kilo\gram} billet (\SI{80x30x210}{\milli\metre}) and hot rolled at \SI{1050}{\celsius} in a 3:1 ratio.
This was followed by an annealing heat treatment of \SI{1250}{\celsius} for \SI{12}{\hour}, then a quench in air to room temperature.
A small dog-bone shape specimen was machined from the billet with a \SI{0.5 x 0.5}{\milli\metre\squared} gauge cross-section and a \SI{2.39}{\milli\metre} gauge length.
For the purposes of this study, the sample was measured in a simple static condition; only the microstructure within the gauge section was of interest.
\subsection{DCT data collection}
\begin{figure}
    \includegraphics{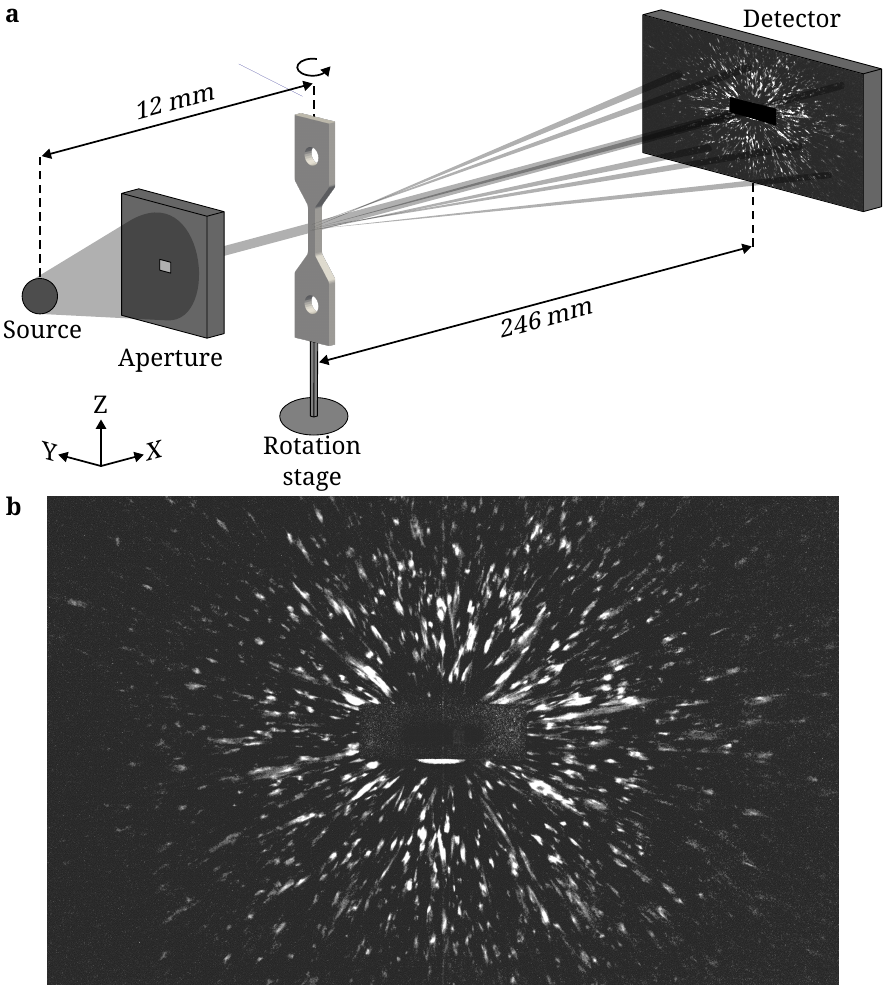}
    \caption{DCT data collection projection geometry (\textbf{a}) and example detector image showing diffracted peaks (\textbf{b}).}
    \label{fig:proj_geom}
\end{figure}

The DCT scans were collected on a ZEISS Xradia 520 Versa X-ray microscope equipped with a LabDCT Pro module and a flat-panel extension.
An accelerating voltage of \SI{110}{\kilo\volt} was used with a power of \SI{10}{\watt}.
A flat-panel detector (\SI{75} {\micro\metre} pixels) was used to collect the diffracted X-ray signal in projection geometry with a source-sample distance of \SI{12}{\milli\metre} and a sample-detector distance of \SI{246}{\milli\metre}, as shown in Figure~\ref{fig:proj_geom}a.
This gives a geometric magnification factor of \num{21.5}.
A \SI{150 x 750}{\micro\metre\squared} beam-defining aperture was placed between the beam and the sample to limit the exposed sample region, while the direct beam was blocked with a beamstop.
A helical phyllotaxis scan strategy \citep{oddershede_advanced_2022} was employed to scan a \SI{\sim 1}{\milli\metre}-tall region of the sample gauge section.
\num{851} projections were captured, see example in Figure~\ref{fig:proj_geom}(\textbf{b}), each with a \SI{60}{\second} exposure time, for a total scan time of 
\SI{16}{\hour}~\SI{45}{\minute}.

\begin{figure}
    \includegraphics{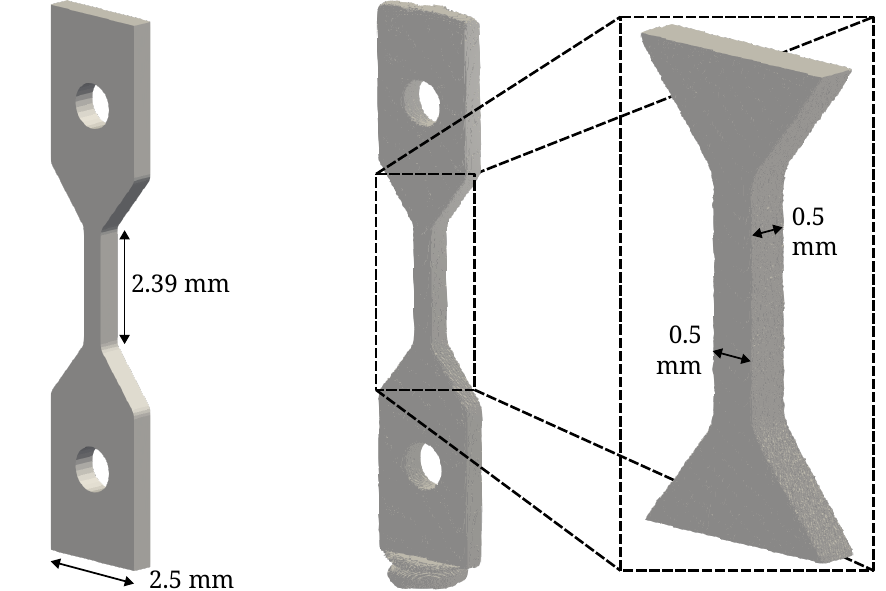}
    \caption{Original sample design (left), \SI{13}{\micro\metre} \emph{coarse} whole-sample tomography reconstruction (centre), \SI{5}{\micro\metre} \emph{fine} gauge-only tomography reconstruction (centre).}
    \label{fig:tomography_recon}
\end{figure}

\num{1601} absorption contrast X-ray tomography (ACT) projections (the \emph{fine} tomography scan) were also taken with a \SI{1}{\second} exposure time and a \SI{5}{\micro\metre} voxel size to define the absorption mask required for the grain reconstruction process.
Finally, a \emph{coarse} whole-sample tomography scan was performed with a \SI{13}{\micro\metre} voxel size, \num{801} ACT projections and a \SI{0.5}{\second} exposure time.
The reconstructed \emph{coarse} and \emph{fine} ACT volumes are presented in Figure~\ref{fig:tomography_recon}.
The reconstruction process utilised to generate the Figure is outlined in \ref{sec:tomo_appendix}.

\subsection{DCT reconstruction}
\begin{figure}
    \includegraphics{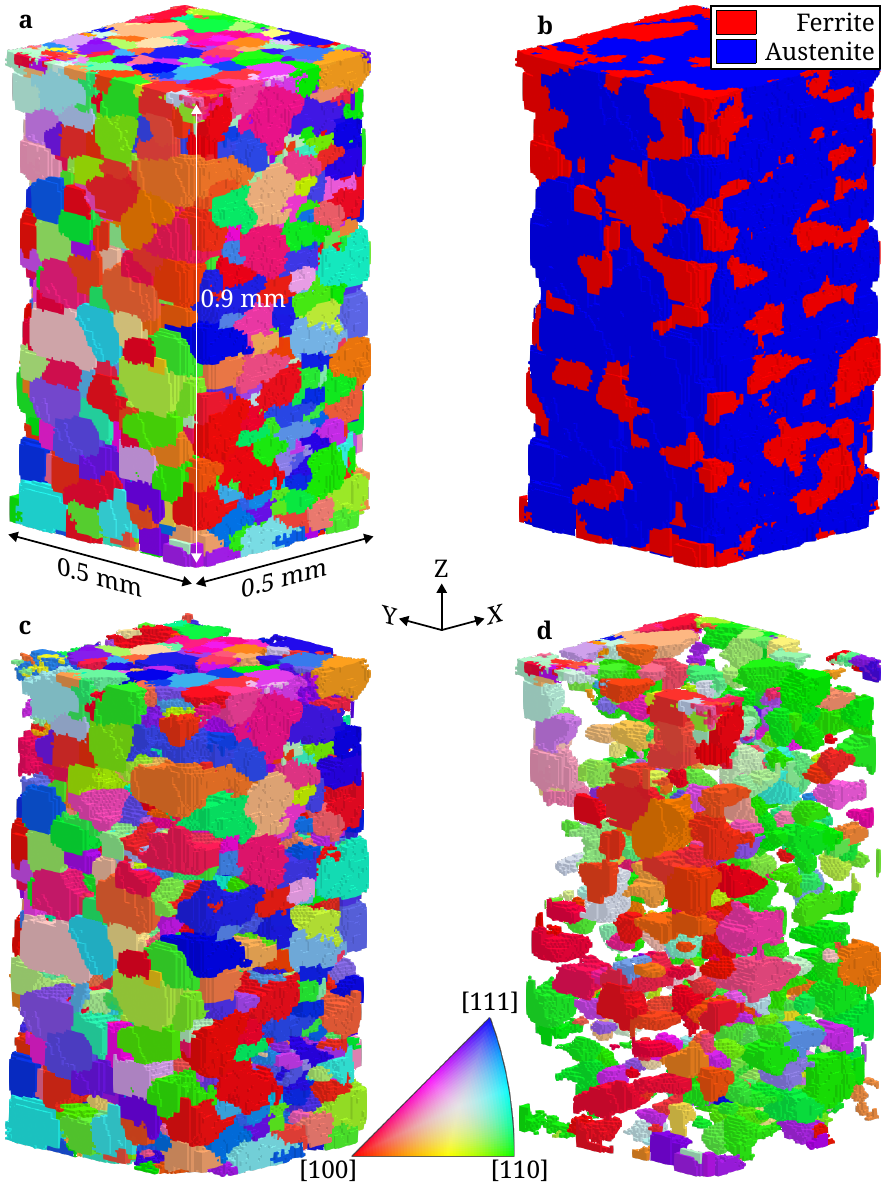}
    \caption{Reconstructed DCT maps. (\textbf{a}, \textbf{b}) whole sample; (\textbf{c}) austenite phase only; (\textbf{d}) ferrite phase only. (\textbf{a}, \textbf{c}, \textbf{d}) are coloured by IPF-$Z$ orientation; (\textbf{b}) is coloured by phase.}
    \label{fig:dct_phase_and_ipf_z_map}
\end{figure}

To reconstruct the final 3D grain map, a prototype version of GrainMapper3D allowing simultaneous indexing of multiple phases was used to process the DCT raw images.
This is an extended version of the fast geometric indexing outlined by \citet{bachmann_3d_2019}, assigning to each voxel in space both the phase and the orientation giving the highest completeness score, where completeness is the ratio between the observed and expected number of reciprocal vectors associated with the solution.
A region of \SI[range-units=single]{660 x 655 x 930}{\micro\metre\cubed} was reconstructed with a \SI{5}{\micro\metre} voxel size, for a total grid of \SI{132 x 131 x 186}{\voxel}.
Grains were defined using a \SI{0.25}{\degree} misorientation threshold between adjacent pixels.
This yielded \num{1888} austenite and \num{685} ferrite grains.
The final result comprised 3D maps of orientation, grain ID, phase ID and completeness, a selection of which are shown in Figure~\ref{fig:dct_phase_and_ipf_z_map}.

\subsection{EBSD data collection}
The sample was mounted in conductive bakelite, polished to a \SI{0.04}{\micro\metre} surface finish using colloidal silica, then electro-polished at \SI{20}{\celsius} with an 80:20 mixture of ethanol and perchloric acid at \SI{15}{\volt} for \SI{20}{\second} with a flow rate of \SI{10}{\litre\per\minute}.
The sample was examined with a JEOL 7000 field emission gun scanning electron microscope (FEG-SEM) equipped with an Oxford Instruments Nordlys EBSD detector to collect an EBSD map across the full width of the sample gauge.
A \SI{1.25}{\micro\metre} step size at a \SI{13}{\nano\ampere} probe current and a \SI{20}{\kilo\electronvolt} beam energy was used.
The EBSD scan and indexing was performed using the Oxford Instruments software AZtec.

\begin{figure*}
    \centering
    \includegraphics{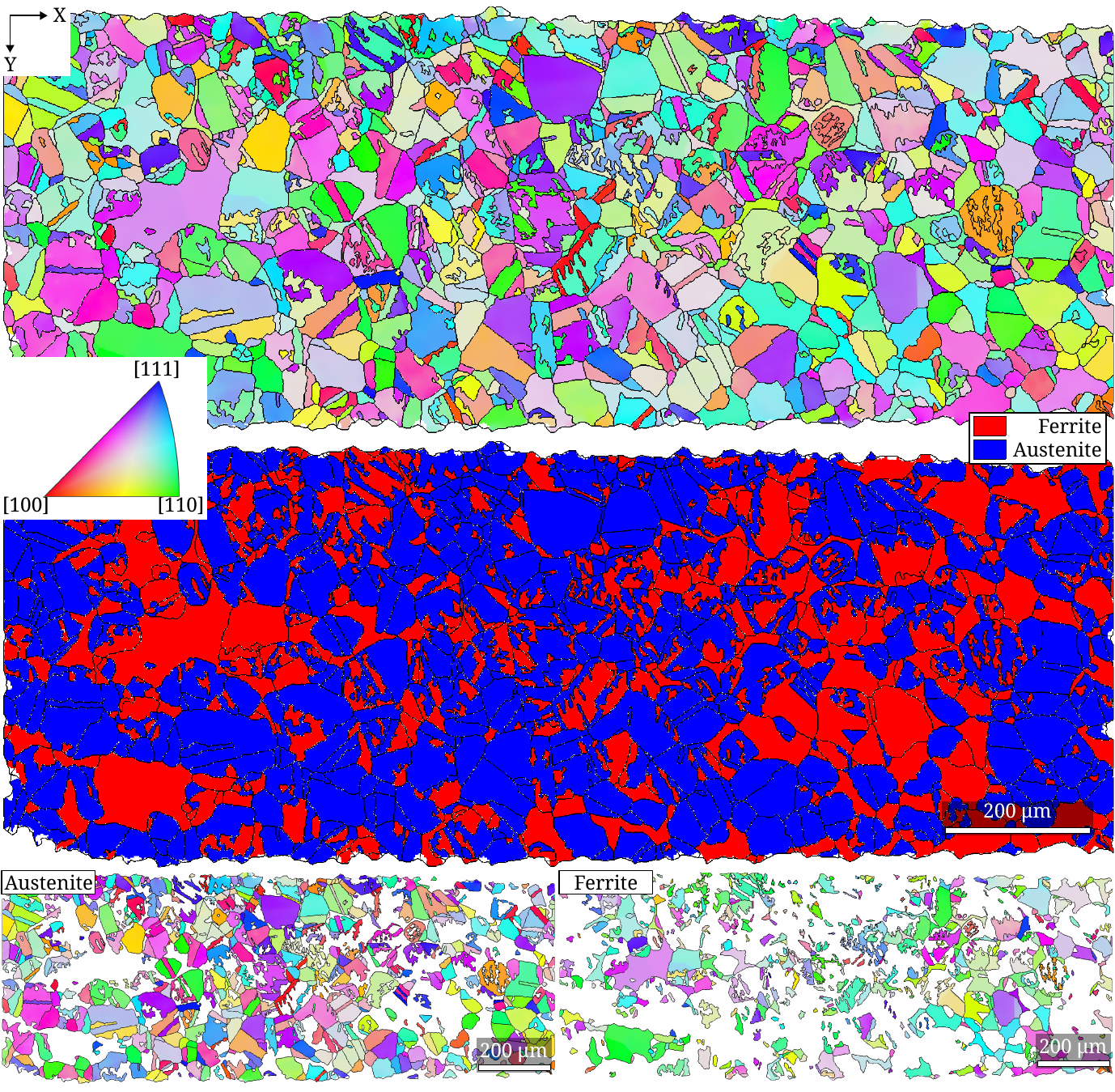}
    \caption{Reconstructed EBSD maps of sample, with IPF-$X$ (axial) orientation colouring (top) and phase colouring (middle). Individual phases are also plotted (bottom) with IPF-$X$ colouring.}
    \label{fig:ebsd_phase_and_ipf_z_map}
\end{figure*}

\subsection{EBSD post-processing}
The EBSD map dataset was imported into the MTEX MATLAB library \citep{bachmann_texture_2010}.
First, the dataset was cropped to the geometry of the sample.
Next, the dataset was segmented into individual grains.
A first pass segmented grains by pixel orientation, with a \SI{5}{\degree} tolerance.
Then, grains with less than 20 contributory pixels were marked as unindexed to exclude grains with potential inaccuracies with grain mean orientation or centroid position.
The first segmentation was then repeated with the updated dataset to regenerate the EBSD grain IDs.
Then, the EBSD map was denoised to fill unindexed pixels within individual grains using an MTEX denoising method with a half-quadratic filter \citep{hielscher_denoising_2019}.
Finally, the grains were re-segmented to re-associate the updated pixels to the grains.
Grain ID and phase ID maps, along with grain-averaged orientations as Euler angles, were exported from MATLAB to an HDF5 file to facilitate further processing with Python.
The processed EBSD map of the sample, comprising \num{750} austenite and \num{648} ferrite grains, is shown in its entirety in Figure~\ref{fig:ebsd_phase_and_ipf_z_map}, with both IPF-$X$ orientation colouring and phase colouring.

\section{Registration}
\subsection{Importing EBSD and DCT data}
A registration procedure was devised to locate the 2D slice within the 3D DCT data that best corresponded to the EBSD image plane.
First, both EBSD and DCT datasets were imported using the pymicro Python library \citep{proudhon_pymicro_2021}, to generate \verb|Microstructure| class instances.
The pymicro library stores grain orientations as a single \num{3x3} orientation matrix per grain ($\matr{g}$) transforming a vector in the sample reference frame ($\overrightarrow{V_s}$) into the crystal reference frame ($\overrightarrow{V_c}$), as per Equation~\ref{eq:pymicro_orien}.
\begin{equation}
    \label{eq:pymicro_orien}
    \overrightarrow{V_c} = \matr{g} \overrightarrow{V_s}
\end{equation}
Due to differences in grain orientation and array axis conventions between GrainMapper3D, MTEX and pymicro, DCT grain ID, grain orientations and phase ID information were verified using the reconstruction report generated by GrainMapper3D, and EBSD grain ID, grain orientations and phase ID information was verified using the MTEX-processed datasets.

\subsection{Initial transformation}
The longitudinal axis of the sample in the original EBSD dataset was parallel to the $X_{E}$ axis of the EBSD reference frame.
In the DCT dataset, the sample longitudinal axis was parallel to the $Z_{D}$ axis of the DCT reference frame.
Consequently, a new rotated EBSD reference frame was devised such that the EBSD sample longitudinal axis was made parallel to the new $Z$ axis ($Z_{R}$).
Given a vector in the original EBSD reference frame ($\overrightarrow{V_E}$), a rotation matrix $\matr{R}$ was defined that transforms the vector into the equivalent vector in the rotated reference frame ($\overrightarrow{V_R}$), as per Equation~\ref{eq:initial_rot}.
\begin{equation}
    \label{eq:initial_rot}
    \overrightarrow{V_R} = \matr{R} \overrightarrow{V_E}
\end{equation}
To represent the EBSD grain orientations ($\matr{g}_{E}$) in the new reference frame, we must right-multiply by the transform of the rotation matrix, as per Equation~\ref{eq:initial_rot_g}.
\begin{equation}
    \label{eq:initial_rot_g}
    \matr{g}_{R} = \matr{g}_{E} \matr{R}^{\intercal}
\end{equation}

\subsection{Initial matching grain search}
Once the EBSD and DCT datasets were approximately aligned by applying this initial transformation, an initial search for matching grain pairs was performed.
A Python function based on the \verb|match_grains| method of the pymicro \verb|Microstructure| library was devised to search for matching EBSD grains within the DCT dataset, as per Algorithm~\ref{alg:get_matches}.
The EBSD microstructure instance was filtered to keep only austenite grains, as initial observations of the crystal orientations revealed a highly textured martensitic-ferrite phase, which may have generated false matches due to grouping of ferrite grains in orientation space.
\num{750} austenite EBSD grains remained after this filtration.
The DCT microstructure instance was similarly filtered, leaving \num{1888} austenite grains.

\begin{algorithm}
\caption{A Python function to find matching grains between EBSD and DCT microstructures.}
\label{alg:get_matches}

\SetKwFunction{FindMatchGrains}{find\_matching\_grains}
\SetStartEndCondition{ }{}{}
\SetKwProg{Fn}{def}{\string:}{}
\SetKw{KwTo}{in}
\SetKwFor{For}{for}{\string:}{}
\SetKwIF{If}{ElseIf}{Else}{if}{:}{elif}{else:}{}

\AlgoDontDisplayBlockMarkers
\SetAlgoNoEnd
\SetAlgoNoLine

\SetKwInOut{Input}{input}
\SetKwInOut{Output}{output}
\SetKwData{gEBSD}{$g_{R}$}
\SetKwData{gDCT}{$g_{DCT}$}
\SetKwData{bestmisorien}{best\_misorien}
\SetKwData{bestmatch}{best\_match}
\SetKwData{EBSDgrain}{EBSD\_grain}
\SetKwData{EBSDindex}{EBSD\_index}
\SetKwData{EBSDgrains}{EBSD\_grains}
\SetKwData{DCTgrain}{DCT\_grain}
\SetKwData{DCTgrains}{DCT\_grains}
\SetKwData{mistol}{mis\_tol}
\SetKwData{misorien}{misorien}
\SetKwData{bestmatches}{best\_matches}
\SetKwFunction{misorientation}{misorientation}
\SetKwFunction{len}{len}
\SetKwFunction{enum}{enumerate}
\SetKwFunction{emptyt}{np.empty}

\newcommand{\forconde}{\EBSDindex, \EBSDgrain \KwTo \enum{\EBSDgrains}}
\newcommand{\forcondd}{\DCTgrain \KwTo\DCTgrains}

\Fn(){\FindMatchGrains{\EBSDgrains, \DCTgrains, \mistol}}{
    \KwData{\\\begin{tabular}{rl}
    \EBSDgrains:& a list of EBSD grains\\
    \DCTgrains:& a list of DCT grains\\
    \mistol:& a tolerance in misorientation (degrees)
            \end{tabular}}
    \KwResult{\\\begin{tabular}{rl}
    \bestmatches:& a list of matched DCT grain IDs\\
                & for each EBSD grain\\
            \end{tabular}}
    \BlankLine
    \tcc{Create empty array with the same length as the number of EBSD grains}
    \bestmatches = \emptyt{\len{\EBSDgrains}}\;
    \tcc{Iterate through input EBSD grains}
    \For{\forconde}{
        \gEBSD = \EBSDgrain rotated orientation matrix\;
        \bestmisorien = \mistol\;
        \bestmatch = -1\;
        \tcc{Iterate through input DCT grains}
        \For{\forcondd}{
            \gDCT = \DCTgrain orientation matrix\;
            \tcc{Use pymicro method to check misorientation between grain orientations}
            \misorien = \misorientation{\gEBSD, \gDCT}\;
            \If{\misorien $<$ \bestmisorien}{
                \bestmisorien = \misorien\;
                \bestmatch = \DCTgrain ID\;
            }
        }
        $\bestmatches[\EBSDindex]$ = \bestmatch\;
    }
}
\end{algorithm}

With a small misorientation tolerance (\SI{3}{\degree}), only \num{222} matching austenite DCT grains were found.
Additionally, the matching DCT grains did not lie on a specific $YZ_{D}$ plane, which would be anticipated for legitimate matches.
It was then theorised that a misorientation remained between the rotated EBSD and DCT reference frames, larger than a \SI{3}{\degree} misorientation tolerance would allow for.
This misorientation was attributed to misalignment between the EBSD spatial and grain orientation reference frames, compounded with spatial distortions introduced by the large field-of-view map collected at low magnifications.
Repeating the search with a wider tolerance of \SI{12}{\degree} also yielded no specific matching plane signal, likely due to the significantly increased noise floor.

\subsection{Corrective rotation search}
Applying a specific corrective rotation to each EBSD grain orientation to realign the EBSD reference frames before searching for matches would negate any remaining misorientation between the datasets and lead to the discovery of legitimate grain matches.
A search through rotation space was therefore required to determine the corrective rotation to apply to each EBSD grain orientation.
To perform the search, a global optimization strategy was employed.
A modified grain matching function was devised, based on Algorithm~\ref{alg:get_matches}.
Before determining the misorientation between grain pairs, $\matr{g}_R$ was right-multiplied by a candidate corrective rotation matrix $\matr{C}$ representing a specific point in rotation space.
Rotation space was parameterized by three successive elemental rotation matrices following the Proper Euler angle ZXZ convention with angles $(\alpha, \beta, \gamma)$ yielding a rotation matrix $\matr{C}$ as per Equation~\ref{eq:zxz}:
\begin{equation}
    \label{eq:zxz}
    \begin{split}
    \matr{C} &= \matr{Z}(\alpha)\matr{X}(\beta)\matr{Z}(\gamma) \\
             &= \begin{bmatrix}
                c_\alpha c_\gamma-c_\beta s_\alpha s_\gamma & -c_\alpha s_\gamma-c_\beta c_\gamma s_\alpha  &  s_\alpha s_\beta\\
                c_\gamma s_\alpha+c_\alpha c_\beta s_\gamma &  c_\alpha c_\beta c_\gamma-s_\alpha s_\gamma  & -c_\alpha s_\beta\\
                s_\beta s_\gamma                            &  c_\gamma s_\beta                             &  c_\beta
                \end{bmatrix}
    \end{split}
\end{equation}
where $s$ and $c$ represent $\sin$ and $\cos$ respectively, and $\matr{X}$ and $\matr{Z}$ represent elemental rotation matrices about fixed-frame axes \citep{roithmayr_dynamics_2015}.
Defining the corrective rotation matrix in this way yields a 3D search space through $(\alpha, \beta, \gamma)$.
A search for grain matches with a tight misorientation tolerance should achieve a maximum number of matches at a specific value of $(\alpha, \beta, \gamma)$, corresponding to the real corrective rotation that must be applied to the EBSD grains in order to bring the EBSD and DCT reference frames into coincidence.

In order to efficiently perform this search through rotation space, the previously defined grain matching algorithm (Algorithm~\ref{alg:get_matches}) was modified to include the three angles of rotation $(\alpha, \beta, \gamma)$ in an array as an input argument, then calculate the corresponding corrective rotation matrix $\matr{C}$, right-multiply the EBSD grain orientation matrix $\matr{g}_R$, and search for matches.
The algorithm was also modified to return only the fraction of missed matches (using the number of matching DCT grains found, and the number of input EBSD grains as the maximum number of potential matches) as a floating-point number between \num{0} and {1}.
This way, the returned number would be equal to \num{1} if no matches were found, and would decrease towards a minimum of \num{0} if all input EBSD grains had a corresponding DCT match.

With an objective function now defined, the PySwarms Python library \citep{miranda_pyswarms_2018} was employed to perform a global minimization over $(\alpha, \beta, \gamma)$ space.
A search space of \SIrange[]{-20}{20}{\degree} in each dimension was employed.
To speed up the evaluation of each point in rotation space, only the \num{10} largest austenite EBSD grains were used.
A small misorientation tolerance of \SI{2}{\degree} was used to minimise the likelihood of the algorithm returning a false positive match.
A global Particle Swarm Optimisation (PSO) based minimisation search was performed with the parameters as per Table~\ref{tab:pso_params}.
The reader is referred to the original definitions of the PSO for further detail on the optimisation parameters \citep{kennedy_particle_1995, shi_modified_1998}.
The search was parallelised across \num{20} cores of an AMD Ryzen 9 3900x CPU.
The global Particle Swarm Optimisation algorithm returned a rotation sequence of $\matr{C}_{\text{PSO}} = \matr{Z}(\SI{-3.23}{\degree})\matr{X}(\SI{9.04}{\degree})\matr{Z}(\SI{3.35}{\degree})$.
\begin{table}[H]
\centering
\caption{Global Particle Swarm Optimisation parameters.}
\label{tab:pso_params}
\begin{tabular}{@{}lc@{}}
\toprule
\textbf{Parameter} & \textbf{Value} \\ \midrule
$c_1$                 & 0.5            \\
$c_2$                 & 0.3            \\
$w$                  & 0.9            \\
$N_{\text{particles}}$        & 20             \\
$N_{\text{iters}}$       & 50            
\end{tabular}
\end{table}

\subsection{Local corrective rotation optimisation}
Once approximate values for $(\alpha, \beta, \gamma)$ were found that maximised the number of matches between EBSD and DCT grains, a local optimisation was performed to further refine these values.
To do this, the original matching algorithm, Algorithm~\ref{alg:get_matches}, was used, still with only the \num{10} largest austenite EBSD grains, to get the matching EBSD and DCT grain ID pairs after applying the optimised rotation $\matr{C}$ to the EBSD grains.
Once the list of matching grain pairs was obtained, a new objective function was obtained that returns the mean misorientation between each grain pair in the matched grains list after applying $\matr{C}$.
This way, more accurate $(\alpha, \beta, \gamma)$ values would result in a reduced mean misorientation.
For the local optimisation search, the \verb|minimize| function from the \verb|scipy.optimize| Python library was used \citep{virtanen_scipy_2020}.
The limited-memory Broyden–Fletcher–Goldfarb–Shanno (L-BFGS) nonlinear optimization method was automatically selected \citep{nocedal_numerical_2006}.
The search space was bounded to \SIrange[]{-4}{4}{\degree} in each dimension around the result determined by the PSO method.
A final optimised rotation of $\matr{C}_{\text{fin}} = \matr{Z}(\SI{-3.46}{\degree})\matr{X}(\SI{10.41}{\degree})\matr{Z}(\SI{3.09}{\degree})$ was determined.

\subsection{DCT slice determination}
\begin{figure}
    \includegraphics{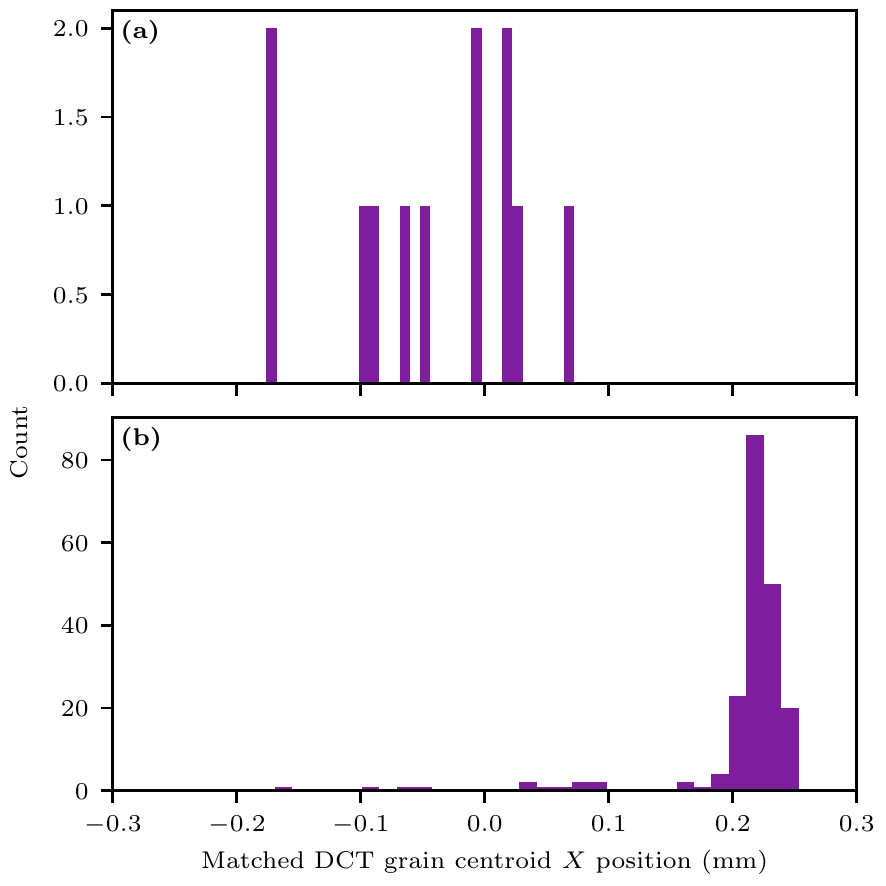}
    \caption{Histograms of $X$ coordinates of centroids of matched DCT grains generated from matching algorithm before (\textbf{a}) and after (\textbf{b}) correcting EBSD orientations.}
    \label{fig:match_histogram}
\end{figure}

Once an optimised corrective rotation matrix was determined, a final search for matching grain pairs was performed, using both ferrite and austenite phases of all DCT and EBSD grains, and a misorientation tolerance of \SI{1}{\degree}.
This match was performed with both the uncorrected and corrected EBSD grain orientations, generating two lists of matched grain pairs.
The results of this grain-matching algorithm were then explored by plotting the $X$ coordinate of the centroids of the DCT grains that were returned by the grain-matching algorithm, both before and after applying the optimized corrective rotation to the input EBSD grains, as shown in Figure~\ref{fig:match_histogram}.

The position in the DCT reference frame of all matched DCT grains in both these lists was then investigated - as the matched DCT grains in the corrected grain pair list were all roughly co-planar with a consistent $X$-axis of their centroid position, the nearest corresponding DCT microstructure slice was extracted and could therefore be compared directly to the transformed EBSD grain map.
From the peak of the histogram in Figure~\ref{fig:match_histogram}\textbf{(b)}, an $X$ slice position of $X=\SI{0.22}{\milli\metre}$ from the centre-of-mass of the DCT scan was determined, corresponding to a 3D array slice index of \num{109}.

\subsection{DCT slice registration}
\begin{figure}
    \centering
    \includegraphics{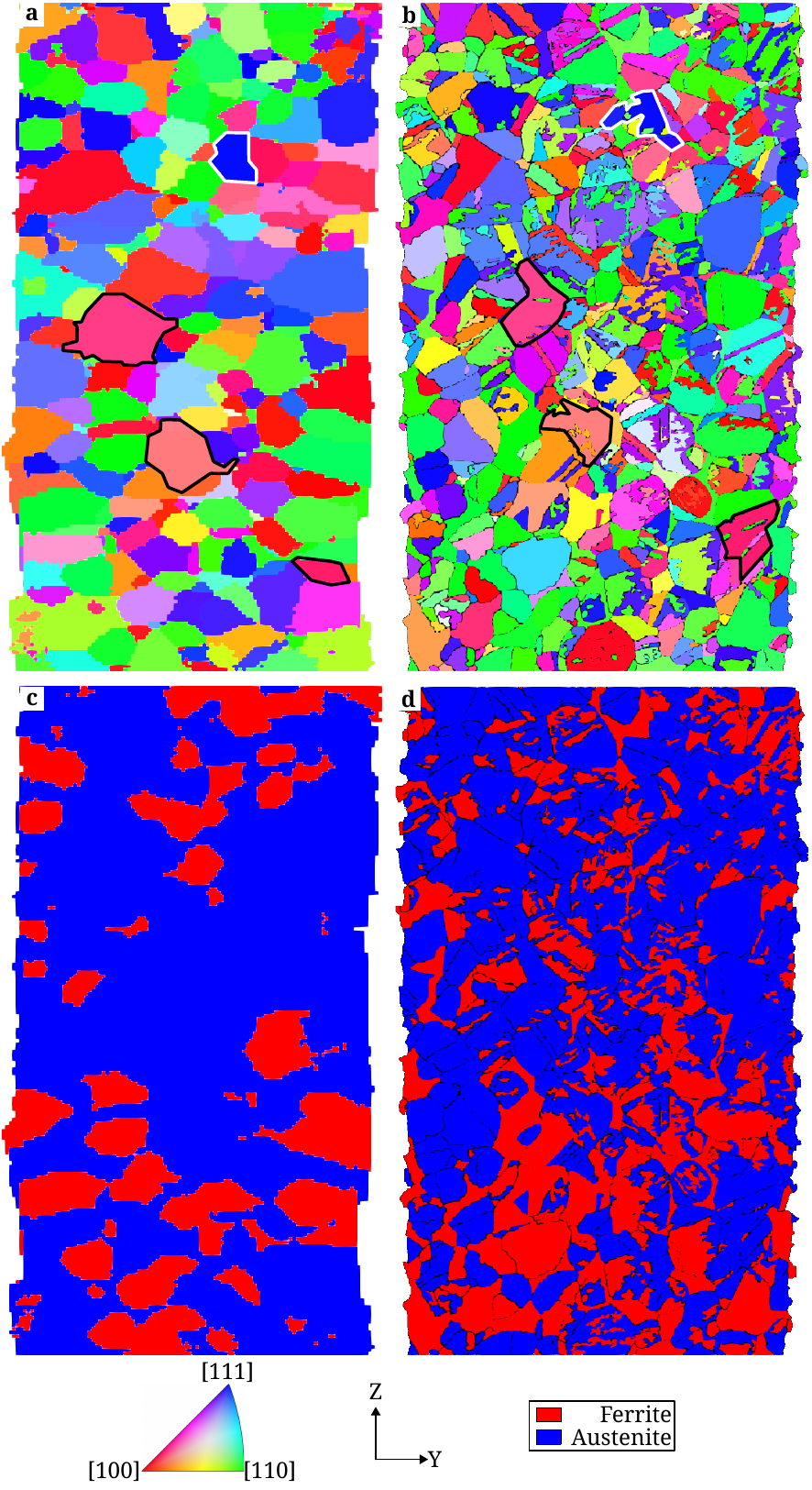}
    \caption{DCT map, YZ slice \num{109} of \num{130} (\textbf{a}, \textbf{c}) vs cropped EBSD map (\textbf{b}, \textbf{d}), coloured by IPF-$Z$ (\textbf{a}, \textbf{b}) and phase (\textbf{c}, \textbf{d}). Select DCT grains are outlined (\textbf{a}) with their corresponding EBSD matches (\textbf{b}).}
    \label{fig:dct_vs_ebsd_ipf}
\end{figure}

As the EBSD map was taken over a larger region of the sample than the DCT map, it was necessary to accurately determine the crop required to generate a new EBSD map representing only the region explored by the DCT scan.
To do this, the 2D centroids of the matching grain pairs were determined in the reference frames of the full EBSD map and the matched DCT slice respectively.
With a list of matching 2D coordinate positions, a rigid transformation was performed to determine the translation required to transform the DCT grain 2D centroids to the EBSD grain centroids.
This translation was then rounded to the nearest EBSD integer pixel, and applied to the origin point of the 2D EBSD map to generate a cropped EBSD sub-region with new grain centroids that closest matched the corresponding DCT centroids.
Figure~\ref{fig:dct_vs_ebsd_ipf} shows the final cropped EBSD map with corrected grain orientations compared to $YZ$ slice \num{109} of the DCT dataset.

\begin{figure*}
    \centering
    \includegraphics{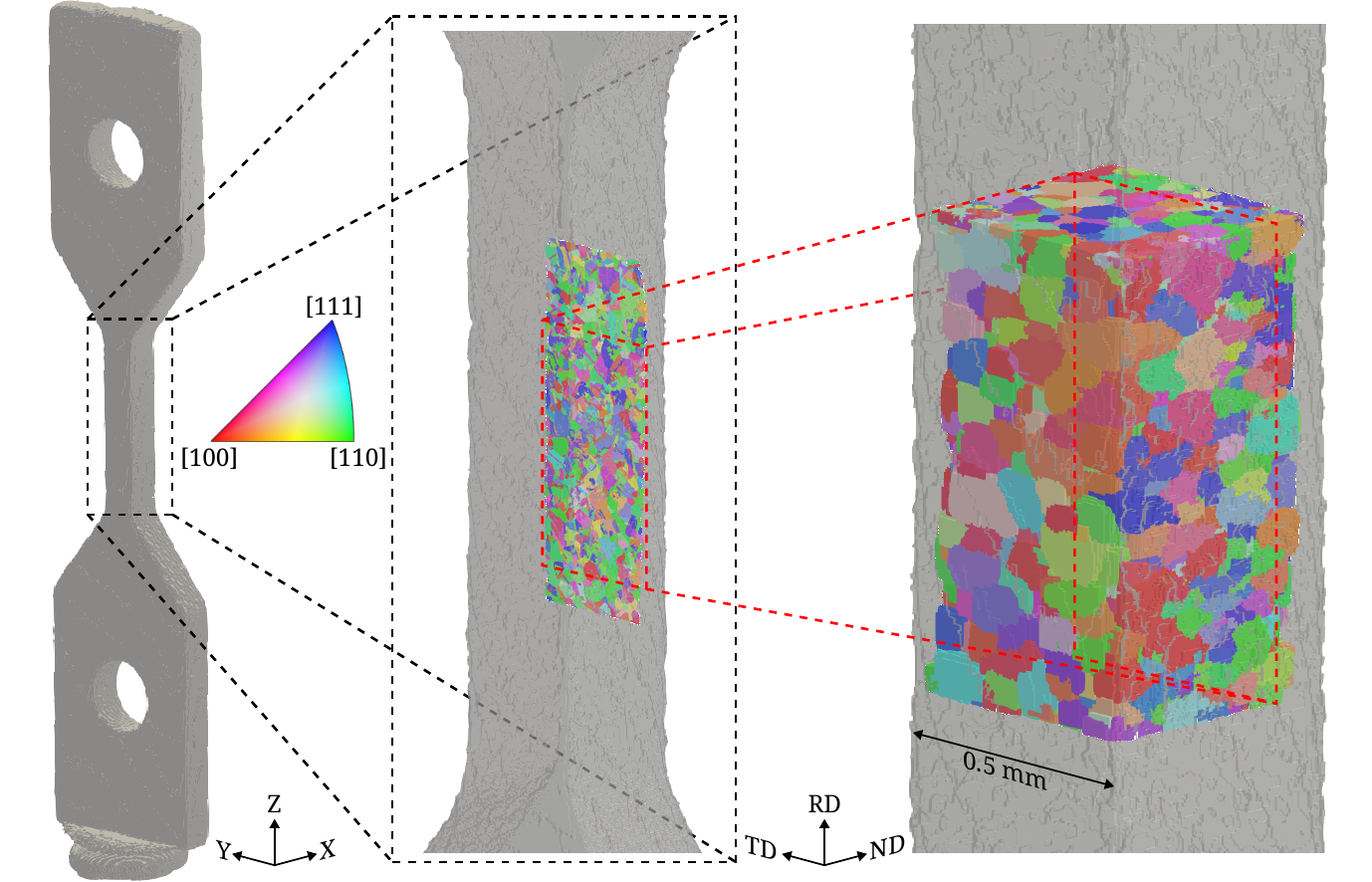}
    \caption{EBSD (centre) and DCT (right) IPF-$Z$ orientation maps embedded in sample geometry from \emph{coarse} tomography (left).}
    \label{fig:post_reg}
\end{figure*}

After limiting matched DCT grains to those that appeared in the $YZ$ slice, a total of \num{166} matching grain pairs were found between the EBSD and DCT grain maps.
With all registrations performed, the embedding of the DCT and EBSD grain maps within the sample geometry (as generated from the tomography data) could then be performed, as presented in Figure~\ref{fig:post_reg}.
Details of the processing pipeline used to generate the tomography model are available in \ref{sec:tomo_appendix}.

\section{Results and Discussion}
With both datasets reconstructed and a registration performed, direct statistical descriptions of and comparisons between the DCT and EBSD grain maps were undertaken.
For all subsequent comparisons, calculations were performed for both the full as-reconstructed EBSD and DCT grain maps (labelled as EBSD (full) and DCT (full) respectively), the EBSD region cropped to the region that overlaps the DCT volume (labeled as EBSD (crop)) as well as the extracted DCT slice.
It is sometimes valuable to distinguish between the DCT 2D slice (similar to an EBSD map, labelled as DCT (2D slice)), and a subset of the 3D DCT dataset that contains only the grains that appeared within the 2D slice (labelled as DCT (3D slice)).

\subsection{DCT 3D grain map}
The reconstructed 3D DCT orientation and phase maps in Figure~\ref{fig:dct_phase_and_ipf_z_map} show an equiaxed microstructure with both austenite and ferrite phases present.
Referring to Figure~\ref{fig:dct_phase_and_ipf_z_map}(\textbf{b}), austenite is the majority phase and ferrite is distributed uniformly throughout the volume.
The grains from the individual phases presented in Figure~\ref{fig:dct_phase_and_ipf_z_map}(\textbf{c}) and (\textbf{d}) show that the volumes of austenite and ferrite are dominated by grains measuring $\sim\SI{100}{\micro\metre}$.
Coloured by the IPF-$Z$ orientation, ferrite possesses distinct bands of similarly oriented grains, as evident with those shown as red ([100] parallel to $Z$) and green ([110] parallel to $Z$).
These macrozones extend along the gauge direction (Z) with a width of three or four grains.

\subsection{EBSD 2D grain map}
The processed 2D EBSD map of the entire sample, shown in Figure~\ref{fig:ebsd_phase_and_ipf_z_map}, displays a dual-phase austenite (approximately equiaxed) and martensitic ferrite (laths) microstructure with the former being the majority phase as in the 3D DCT grain map.
Given the higher spatial resolution of the EBSD measurements over DCT, the grain morphologies are more accurately determined.
Features such as twins, for example, are clearly evident in the austenite phase in the EBSD map.
Similar to the DCT, the ferrite has an interconnected network of grains, as is evident by several neighbouring grains of the same phase.
The macrozone feature, as was very clear in the 3D DCT reconstruction, is not replicated in the EBSD 2D slice.
It is further evident from the EBSD that the grain shape for both phases is complex; this aspect will be quantified in a later section.

\subsection{Phase balance}
\begin{figure}
    \centering
    \includegraphics{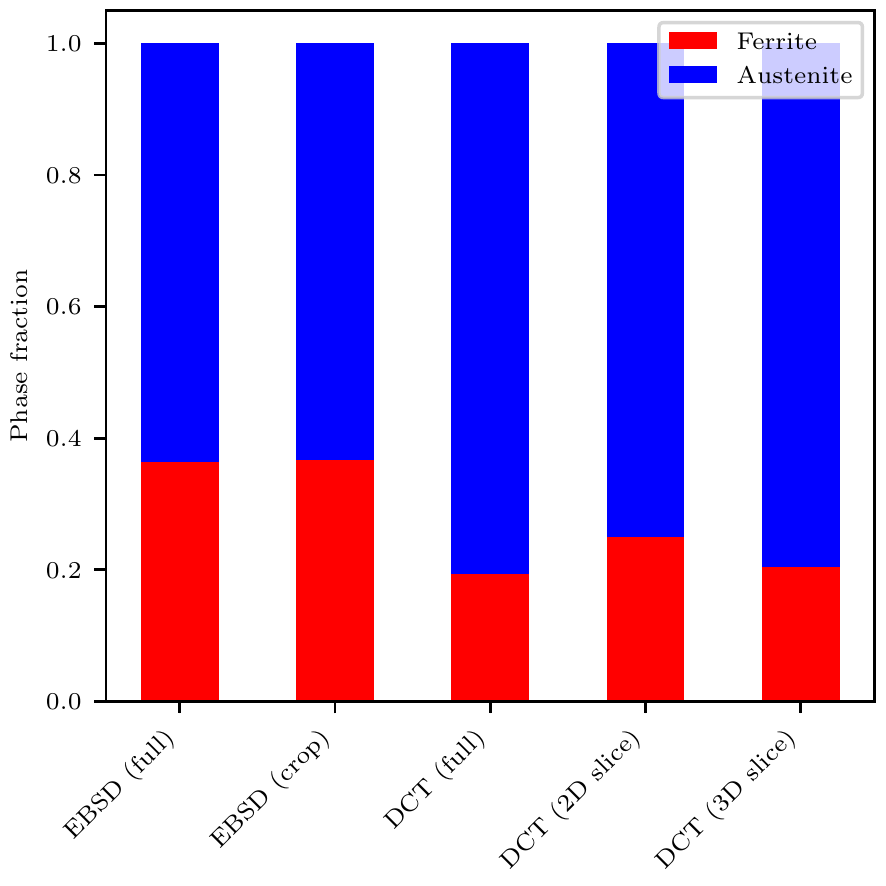}
    \caption{Phase balance comparisons between EBSD (full), EBSD (cropped), DCT (full), DCT (2D slice) and DCT (3D slice).}
    \label{fig:phase_balance}
\end{figure}

Figure~\ref{fig:phase_balance} shows the phase balance comparisons, calculated by pixel area or volume fraction, for both EBSD and DCT datasets.
A significant difference in per-voxel phase fraction is observed between the EBSD and DCT grain maps.
The EBSD map suggests a microstructure that is approximately \SI{40}{\percent} ferritic by area, but DCT datasets average only around \SI{20}-\SI{25}{\percent} ferritic by area or volume.

The larger austenite volume fraction in the DCT experiment may be related to the lattice distortions caused by residual stressed in the austenite grains, which make the diffraction spots from this phase streak radially (see Figure~\ref{fig:proj_geom}(\textbf{b})) leading to overestimated austenite grain sizes.
Furthermore, according to the high resolution EBSD map (see Figure~\ref{fig:ebsd_phase_and_ipf_z_map} and the corresponding grain size distributions in Figure~\ref{fig:grain_size_comparison}) the ferritic grains tend to be smaller than the austenitic grains.
In fact most of these are so small that they fall below the detection limit of the DCT method, which is of the order \SIrange[range-phrase=--, range-units = single]{10}{40}{\micro\metre} \citep{oddershede_non-destructive_2019} for undeformed samples with well defined diffraction spots displaying no radial streaking, hence probably somewhat poorer for the present sample.
Thus it is possible that a large number of smaller ferrite grains were just missed in the DCT experiment, and their corresponding voxels assigned to nearby austenite grains during the reconstruction, creating the observed phase imbalance.
To additionally exclude the contribution of small surface grains being missed by the DCT reconstruction (and therefore leading to larger-than-expected grains at the edges of the sample), the same phase balance calculation was performed for a \emph{trimmed} DCT dataset with \SI{20}{\micro\metre} of data trimmed in the $X$ and $Y$ directions -- this had little effect on the phase balance results, thereby excluding surface effects as a primary cause of the differences between DCT and EBSD techniques.

\subsection{Grain size}
\begin{figure}
    \centering
    \includegraphics{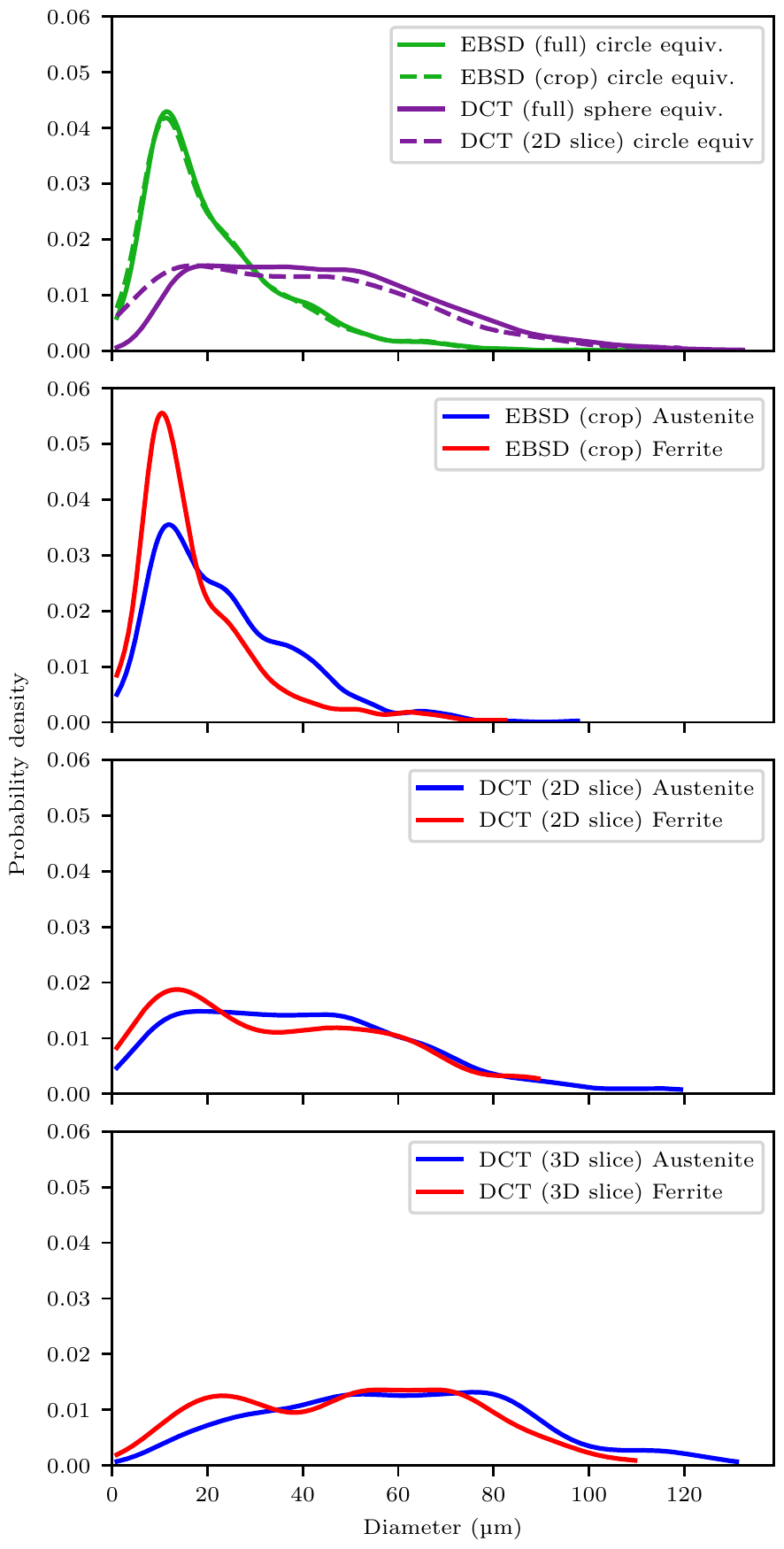}
    \caption{Grain size distributions from EBSD and DCT datasets.}
    \label{fig:grain_size_comparison}
\end{figure}

Figure~\ref{fig:grain_size_comparison} shows grain size distributions of the full EBSD and DCT datasets, as well as the extracted DCT slice, and  phase-specific grain diameter distributions from EBSD (crop) and DCT (2D and 3D slice) datasets.
Both 2D DCT grain circle-equivalent diameters (as estimated from only the 2D DCT slice) and 3D sphere-equivalent diameters (as estimated from all 3D DCT grains that appear in the 2D slice) are provided.

It is evident that there is a significant difference in grain size presentation between DCT and EBSD techniques.
EBSD grains, both in the full and cropped grain maps, possess significantly smaller diameters overall, with a peak diameter of around \SI{15}{\micro\metre}.
EBSD grain diameters have an approximately log-normal distribution.
In contrast, DCT grains present substantially larger on average, with a larger distribution of sizes, and a peak in the full dataset of approximately \SI{40}{\micro\metre}.
Additionally, there were very few EBSD grains observed with diameters \SI{\geq 60}{\micro\metre}, but a significant fraction of DCT grains in the full dataset have diameters larger than this.
To ensure that this effect was not solely caused by the size underestimation common in 2D sections of 3D volumes \citep{astm_international_standard_2021}, DCT grain diameters were calculated for the full 3D volume as well as how they appear in the 2D slice.

Regarding the phase specific grain size distributions, ferrite grains appear smaller than austenite grains in the cropped EBSD map.
The reasons for this difference between phases being much less pronounced for the DCT map, both for 2D and 3D slices, are undoubtedly the influence of lattice distortion in the austenite phase and the grain size detection limit of the DCT technique, as described above for the phase fractions.
A challenging grain morphology is also a likely reason for DCT grain appearing larger in size as outlined further in the following section.

\subsection{Grain shape}
\begin{figure}
    \centering
    \includegraphics{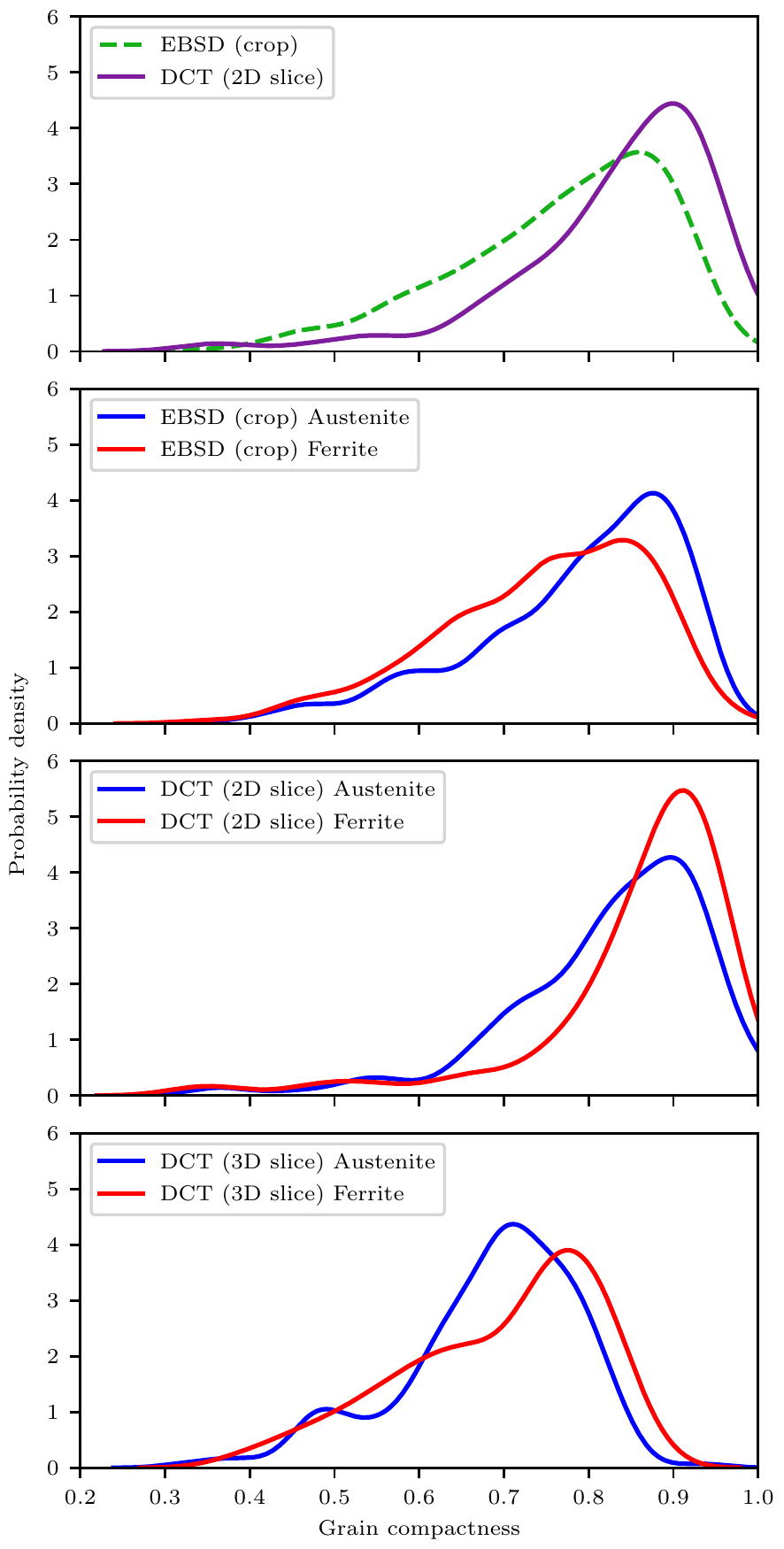}
    \caption{Grain compactness distributions for EBSD and DCT datasets.}
    \label{fig:grain_shape_comparison}
\end{figure}

The compactness, as calculated by taking the ratio between the grain area (or volume) and the area (or volume) of the complex hull of the grain, provides a metric for grain morphology, with values close to 1 representing smoother, more circular grains.
After determining grain size distributions for Figure~\ref{fig:grain_size_comparison}, grains consisting of fewer than \num{10} pixels (\SI{12.5}{\micro\metre} for the EBSD dataset, \SI{50}{\micro\metre} for the DCT dataset) were removed from each dataset prior to compactness calculations to avoid non-physical convex hull results.
Figure~\ref{fig:grain_shape_comparison} shows phase-specific grain compactness distributions from EBSD (crop) and DCT (2D and 3D slice) grain maps.

It is evident from the EBSD map that a large number of fairly complex grain boundaries are present, primarily between grains with different phases, which is common for duplex steels \citep{yousefian_microstructure_2021, zhang_three-dimensional_2019}.
Quantifying the phase specific grain shapes of the EBSD map in Figure~\ref{fig:grain_shape_comparison}, the austenite grains appear slightly more compact than the ferrite grains, indicating a simpler austenite grain morphology.
This is to be expected due to the thermal history of the sample - the austenite grain morphology is inherited from the annealing stage, but the ferrite grew rapidly during the quenching stage.

Interestingly, the DCT reconstruction appears to capture the same compactness distribution for the austenite grains in the 2D slice as for the EBSD map.
This is likely because the ferrite laths take up only very small area fractions of the austenite convex hulls, and with this approximation the 2D shapes of the austenite grains appear similar between EBSD and DCT.
Contrary to this, the DCT 3D slice compactness of the austenite grains is significantly different.
This marked difference in compactness between DCT 2D and 3D slice highlights the need for 3D characterisation techniques (rather than traditional 2D techniques such as EBSD) in order to accurately capture the complicated grain morphology present in this alloy system.

The ferrite grains appear more compact and with a much narrower distribution of compactness than the austenite grains in the DCT 2D and 3D slices, contrary to the observations made for the EBSD map.
The aforementioned complex grain boundaries between phases observed in the EBSD map clearly pose a challenge for the DCT reconstruction - many of the intricate features at the edge of the grain boundaries are only a few pixels in size, and, owing to the differences in technique resolution, are poorly represented in the DCT slice.
More accurate grain morphologies have been observed in prior lab-based DCT studies \citep{king_first_2013, mcdonald_non-destructive_2015, eguchi_x-ray_2022}, and there is no reason to suspect that the challenges encountered here are related to the new dual phase indexing algorithm rather than the complexity of the sample microstructure.

\subsection{Texture}
\begin{figure}
    \centering
    \includegraphics{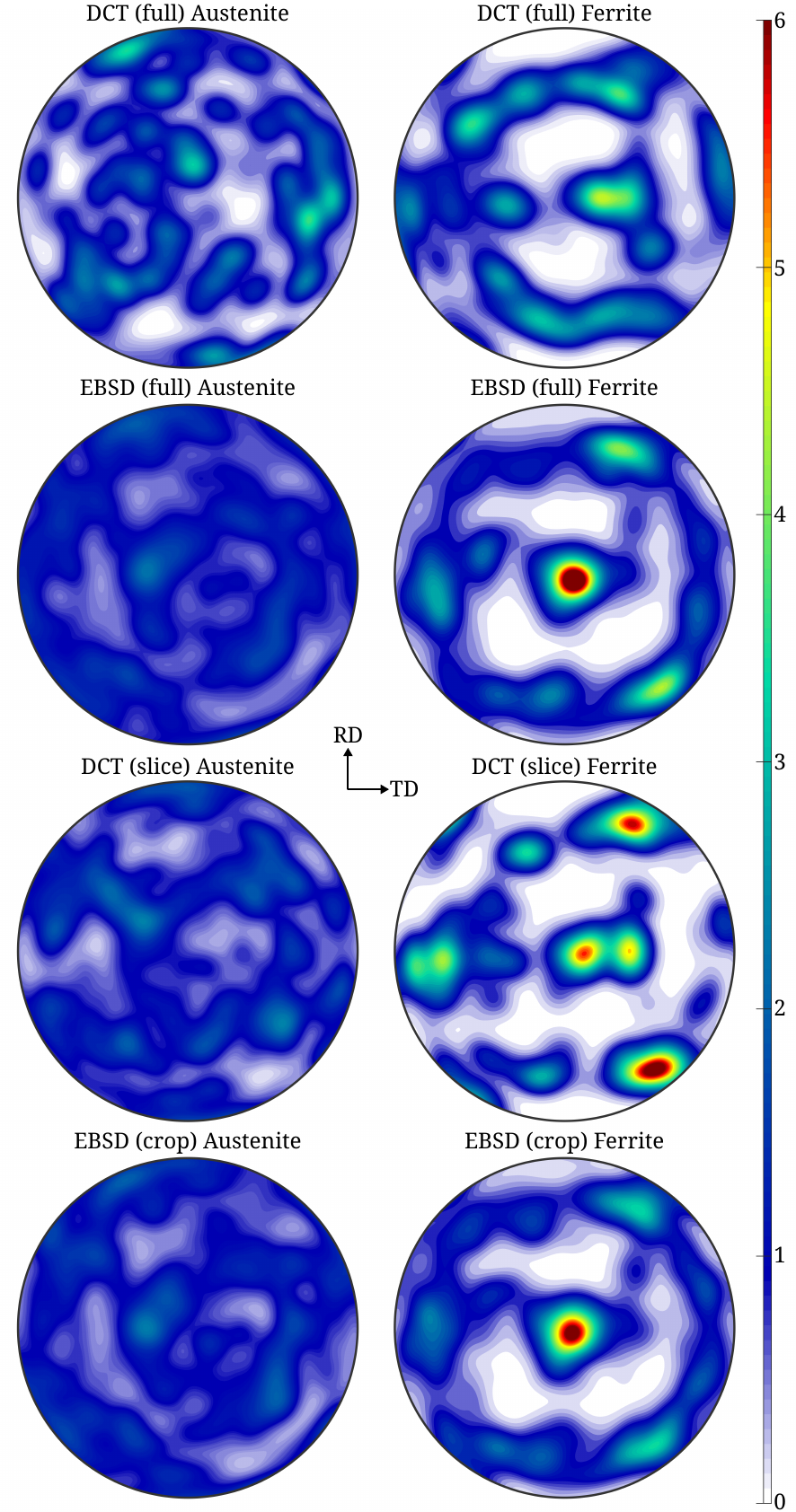}
    \caption{\hkl{111} orientation distribution functions from all EBSD and DCT grains.}
    \label{fig:texture_comparison}
\end{figure}

Figure~\ref{fig:texture_comparison} shows comparisons in austenite and ferrite texture between EBSD and DCT grain maps.
It is clear that the austenite phase is broadly untextured in both the DCT and EBSD datasets.
A substantial \hkl{111}\hkl<112> texture is observed in the ferrite phase in the EBSD (full) scan in Figure~\ref{fig:texture_comparison}, typical for rolled body-centered cubic materials \citep{fu_general_2021}.
Interestingly, a clear difference between the ferrite texture is observed between EBSD and DCT techniques.
This may be caused by a more complicated 3D texture which the DCT (full) plot is capturing - a close look at Figure~\ref{fig:dct_phase_and_ipf_z_map}\textbf{d} supports this hypothesis as large regions of the DCT volume have broadly different ferrite textures.
This explains the more complicated presentation of texture visible in the DCT (full) Ferrite subplot of Figure~\ref{fig:texture_comparison}.
However, when grains being plotted are filtered to just those in the registered slice, as shown in the DCT (slice) Ferrite subplot, the result matches much closer with the EBSD (crop) Ferrite subplot.
This indicates that the DCT technique is accurately determining sample texture of both phases, but the more complicated 3D texture of the ferrite phase is poorly captured by a single 2D slice, demonstrating the need for 3D characterisations to accurately determine sample texture.

\subsection{Registration}
Figure~\ref{fig:dct_vs_ebsd_ipf} demonstrates the validity of the proposed registration technique to determine the location of the 2D slice in the DCT volume that best corresponds to the EBSD map.
A substantial number of larger grains in the cropped EBSD dataset can be identified within the DCT 2D slice.
Grain orientations as represented by their IPF colour appear well matched between techniques, and many matching grain pairs can be visually identified solely by their similar centre-of-mass position and IPF colour.

The proposed registration algorithm has a number of advantages when compared to other registration techniques defined in the literature.
\textit{PolyProc}, for example \citep{kang_polyproc_2019}, includes an advanced registration technique that utilises genetic algorithms to iteratively align two 3D datasets by maximising the number of shared voxels between datasets using the shape of the sample.
This involves a search through 6D space as both translations and rotations are explored.
Due to the increase in efficiency from the parallelised genetic algorithm search, the \textit{PolyProc} registration technique is relatively fast, taking approximately \SI{10}{\minute} to align two 3D datasets.
\textit{PolyProc} is well-suited to alignment of 3D datasets, but can only be applied in the case where the input datasets are both 3D and recorded with the same apparatus.

The registration technique employed by \citet{renversade_comparison_2016} minimises the mean misorientation between corresponding voxels in two different datasets by modifying the magnification, distortion, shear, rotation and translation, leading to a 9D search, which they performed using non-linear optimisation.
This is a robust technique and was successfully applied to register a 2D EBSD slice within a DCT volume.
However, the technique is likely significantly slower to apply than the \textit{PolyProc} technique, due to the high computational cost of misorientation calculation between a large number of pixels, although the problem is easily parallelised by splitting the misorientation calculations across different processor cores.

In contrast, the technique applied in this paper relies only on the average crystallographic orientation of grains (a 3D search) which makes it highly applicable to a range of materials characterisation techniques such as 3DXRD, EBSD, and HEDM.
As a subset of \num{10} grains were selected for both the parallel PSO search and final L-BFGS search, this registration technique is fast, with the entire slice determination procedure taking less than \SI{5}{\minute}.

\subsection{Outlook}

First established at the European Synchrotron Radiation Facility in 2008 \cite{ludwig_x-ray_2008,johnson_x-ray_2008}, then subsequently implemented at other synchrotrons such as SPring-8 \citep{shiozawa_4d_2016} and Soleil \citep{ribart_situ_2023}, synchrotron-based DCT has proved to be a excellent tool for revealing metallurgical phenomena related to deformation and microstructure.
Whilst DCT performed at a synchrotron facility has a number of benefits; foremost is the high-intensity of the X-ray sources, offering rapid acquisition rates necessary for time-resolved studies, difficulties include significant lead times and set-up costs, limited availability and accessibility restrictions.
Such constraints are alleviated with lab-based DCT results; initially developed in 2013 \citep{king_first_2013} and later made commercially available on select ZEISS X-ray microscopes under the name LabDCT™ \citep{mcdonald_non-destructive_2015, holzner_diffraction_2016}. 
Improvements in grain shape reconstructions \citep{bachmann_3d_2019}, acquisition strategies \citep{oddershede_advanced_2022, ganju_novel_2023} and the data reconstruction pipeline \citep{sun_recent_2022} have been significant.
If one considers synchrotron-based DCT as the benchmark to which lab-based DCT will target, future advancements must collectively consider developments to hardware, data collection and post-processing.

In this study, several differences have been identified between the characterisations performed by EBSD and DCT techniques.
The complicated grain morphologies caused by the dual-phase microstructure, combined with lattice distortions in one of the phases and the presence of a number of small grains in the other, presented significant challenges for the DCT grain reconstruction process.
However, important phase specific features were observable including a subsurface ferrite grain network of similarly oriented grains; an observation that was absent in the EBSD map.
It is suggested that the technique is well suited for the observation of multi-phase systems, particularly for well chosen microstructures with coarse grains of simple morphologies.
There is no reason why in-situ lab-based DCT experiments cannot be performed on such multi-phase materials, particularly for experiments that are unfeasible at synchrotrons (e.g. extended duration experiments, high-risk pilot studies, or experiments that demand a rapid lead-time from conception to results).
There is significant promise that new science will be possible using the lab-based DCT method on a compendium of multi-phase, polycrystalline materials across several disciplines.

\section{Conclusions}
In this study a dual-phase steel sample was analysed using two techniques - DCT and EBSD.
A novel registration technique was developed and used to identify the 2D slice within the 3D DCT volume that best corresponds to the EBSD grain map.
Many larger EBSD and DCT grains with matching IPF colours were visually identified, demonstrating the success of the registration algorithm in determining the correct 2D DCT slice.
The algorithm is fast, and requires only grain-averaged orientations, so could therefore be adopted by similar techniques such as 3DXRD and HEDM.

From the EBSD scan, a complicated grain morphology was observed, especially at the interfaces between the austenitic and ferritic phases where a lath-like morphology was present with characteristic lengths on the order of \SIrange[range-phrase=--, range-units = single]{10}{20}{\micro\metre}.
These fine microstructural details were similar in size to the size detection limit of the DCT technique, which influenced a number of differences in microstructural statistics such as phase fractions, grain size and shape distributions, as outlined:
\begin{enumerate}
    \item{The sample was measured as \SI{40}{\percent} ferritic by area using EBSD data. DCT techniques showed a ferrite phase fraction of \SI{20}{\percent} both by area and volume. This discrepancy is ascribed to differences in austenite and ferrite grain size - smaller ferrite grains may have been missed by the DCT reconstruction and incorrectly assigned to neighboring austenite grains, skewing the phase balance results.}

    \item{Grain diameters measured with EBSD appeared significantly smaller than diameters measured with DCT. Furthermore, a difference in grain diameter distributions was observed between ferrite and austenite phases in the EBSD scan, with ferrite grains appearing smaller in diameter. This difference was also observed in the DCT datasets but to a much lesser degree.}

    \item{In the EBSD dataset, ferrite grains were less spherical than austenite grains, whereas the DCT dataset reports ferrite grains as more spherical.}

\end{enumerate}
    
While the austenite phase was broadly untextured in both EBSD and DCT scans, a strong \hkl{111}\hkl<112> rolling texture was observed in the ferrite phase with EBSD.
In contrast, a more complicated texture was observed in the ferrite phase with DCT.
This difference was resolved when plotting the texture of just the grains in the 2D DCT slice.
This indicates a complex 3D texture of columnar networks of similarly orientated ferrite grains in the sample; an important microstructural feature absent from the 2D measurements, demonstrating the need for 3D characterisations to accurately determine texture.

\section{Datasets and Code Access}
Example datasets and analysis code used in this study are available upon request.

\section{Acknowledgments}
James Ball, Stefan Michalik, and David Collins acknowledge the financial support from the Diamond Light Source and the University of Birmingham for a PhD studentship. Claire Davis and Carl Slater acknowledge the Engineering and Physical Sciences Research Council for support via grants EP/P020755/01 and EP/V007548/1.

\clearpage

\newpage
\typeout{}
\bibliography{references.bib}

\begin{thebibliography}{53}
\expandafter\ifx\csname natexlab\endcsname\relax\def\natexlab#1{#1}\fi
\providecommand{\url}[1]{\texttt{#1}}
\providecommand{\href}[2]{#2}
\providecommand{\path}[1]{#1}
\providecommand{\DOIprefix}{doi:}
\providecommand{\ArXivprefix}{arXiv:}
\providecommand{\URLprefix}{URL: }
\providecommand{\Pubmedprefix}{pmid:}
\providecommand{\doi}[1]{\href{http://dx.doi.org/#1}{\path{#1}}}
\providecommand{\Pubmed}[1]{\href{pmid:#1}{\path{#1}}}
\providecommand{\bibinfo}[2]{#2}
\ifx\xfnm\relax \def\xfnm[#1]{\unskip,\space#1}\fi
%Type = Article
\bibitem[{Sedm\'ak et~al.(2016)Sedm\'ak, Pilch, Heller, Kope\v{c}ek, Wright,
  Sedl\'ak, Frost, and \v{S}ittner}]{sedmak2016}
\bibinfo{author}{P.~Sedm\'ak}, \bibinfo{author}{J.~Pilch},
  \bibinfo{author}{L.~Heller}, \bibinfo{author}{J.~Kope\v{c}ek},
  \bibinfo{author}{J.~Wright}, \bibinfo{author}{P.~Sedl\'ak},
  \bibinfo{author}{M.~Frost}, \bibinfo{author}{P.~\v{S}ittner},
\newblock \bibinfo{title}{Grain-resolved analysis of localized deformation in
  nickel-titanium wire under tensile load},
\newblock \bibinfo{journal}{Science} \bibinfo{volume}{353}
  (\bibinfo{year}{2016}) \bibinfo{pages}{559--562}.
%Type = Article
\bibitem[{Nair et~al.(2019)Nair, Nygren, and Pagan}]{Nair2019}
\bibinfo{author}{S.~Nair}, \bibinfo{author}{K.~Nygren},
  \bibinfo{author}{D.~Pagan},
\newblock \bibinfo{title}{{Micromechanical Response of Crystalline Phases in
  Alternate Cementitious Materials using 3-Dimensional X-ray Techniques}},
\newblock \bibinfo{journal}{Scientific Reports} \bibinfo{volume}{9}
  (\bibinfo{year}{2019}) \bibinfo{pages}{18456}.
%Type = Article
\bibitem[{Rodek et~al.(2007)Rodek, Poulsen, Knudsen, and Herman}]{Rodek2007}
\bibinfo{author}{L.~Rodek}, \bibinfo{author}{H.~F. Poulsen},
  \bibinfo{author}{E.~Knudsen}, \bibinfo{author}{G.~T. Herman},
\newblock \bibinfo{title}{{A stochastic algorithm for reconstruction of grain
  maps of moderately deformed specimens based on X-ray diffraction}},
\newblock \bibinfo{journal}{Journal of Applied Crystallography}
  \bibinfo{volume}{40} (\bibinfo{year}{2007}) \bibinfo{pages}{313 -- 321}.
%Type = Article
\bibitem[{Menasche et~al.(2020)Menasche, Shade, and Suter}]{Menasche2020}
\bibinfo{author}{D.~B. Menasche}, \bibinfo{author}{P.~A. Shade},
  \bibinfo{author}{R.~M. Suter},
\newblock \bibinfo{title}{{Accuracy and precision of near-field high-energy
  diffraction microscopy forward-model-based microstructure reconstructions}},
\newblock \bibinfo{journal}{Journal of Applied Crystallography}
  \bibinfo{volume}{53} (\bibinfo{year}{2020}) \bibinfo{pages}{107--116}.
%Type = Article
\bibitem[{Shen et~al.(2020)Shen, Liu, and Suter}]{SHEN2020100852}
\bibinfo{author}{Y.-F. Shen}, \bibinfo{author}{H.~Liu}, \bibinfo{author}{R.~M.
  Suter},
\newblock \bibinfo{title}{{Voxel-based strain tensors from near-field High
  Energy Diffraction Microscopy}},
\newblock \bibinfo{journal}{Current Opinion in Solid State and Materials
  Science} \bibinfo{volume}{24} (\bibinfo{year}{2020}) \bibinfo{pages}{100852}.
%Type = Article
\bibitem[{Li and Suter(2013)}]{Li2013512}
\bibinfo{author}{S.~Li}, \bibinfo{author}{R.~Suter},
\newblock \bibinfo{title}{Adaptive reconstruction method for three-dimensional
  orientation imaging},
\newblock \bibinfo{journal}{Journal of Applied Crystallography}
  \bibinfo{volume}{46} (\bibinfo{year}{2013}) \bibinfo{pages}{512 – 524}.
%Type = Article
\bibitem[{Ludwig et~al.(2008)Ludwig, Schmidt, Lauridsen, and
  Poulsen}]{ludwig_x-ray_2008}
\bibinfo{author}{W.~Ludwig}, \bibinfo{author}{S.~Schmidt},
  \bibinfo{author}{E.~M. Lauridsen}, \bibinfo{author}{H.~F. Poulsen},
\newblock \bibinfo{title}{X-ray diffraction contrast tomography: a novel
  technique for three-dimensional grain mapping of polycrystals. {I}. {Direct}
  beam case},
\newblock \bibinfo{journal}{Journal of Applied Crystallography}
  \bibinfo{volume}{41} (\bibinfo{year}{2008}) \bibinfo{pages}{302--309}.
%Type = Article
\bibitem[{Johnson et~al.(2008)Johnson, King, Honnicke, Marrow, and
  Ludwig}]{johnson_x-ray_2008}
\bibinfo{author}{G.~Johnson}, \bibinfo{author}{A.~King}, \bibinfo{author}{M.~G.
  Honnicke}, \bibinfo{author}{J.~Marrow}, \bibinfo{author}{W.~Ludwig},
\newblock \bibinfo{title}{X-ray diffraction contrast tomography: a novel
  technique for three-dimensional grain mapping of polycrystals. {II}. {The}
  combined case},
\newblock \bibinfo{journal}{Journal of Applied Crystallography}
  \bibinfo{volume}{41} (\bibinfo{year}{2008}) \bibinfo{pages}{310--318}.
%Type = Article
\bibitem[{Shiozawa et~al.(2016)Shiozawa, Nakai, Miura, Masada, Matsuda, and
  Nakao}]{shiozawa_4d_2016}
\bibinfo{author}{D.~Shiozawa}, \bibinfo{author}{Y.~Nakai},
  \bibinfo{author}{R.~Miura}, \bibinfo{author}{N.~Masada},
  \bibinfo{author}{S.~Matsuda}, \bibinfo{author}{R.~Nakao},
\newblock \bibinfo{title}{{4D} evaluation of grain shape and fatigue damage of
  individual grains in polycrystalline alloys by diffraction contrast
  tomography using ultrabright synchrotron radiation},
\newblock \bibinfo{journal}{International Journal of Fatigue}
  \bibinfo{volume}{82} (\bibinfo{year}{2016}) \bibinfo{pages}{247--255}.
%Type = Article
\bibitem[{Ribart et~al.(2023)Ribart, King, Ludwig, Bertoldo, and
  Proudhon}]{ribart_situ_2023}
\bibinfo{author}{C.~Ribart}, \bibinfo{author}{A.~King},
  \bibinfo{author}{W.~Ludwig}, \bibinfo{author}{J.~P.~C. Bertoldo},
  \bibinfo{author}{H.~Proudhon},
\newblock \bibinfo{title}{In situ synchrotron {X}-ray multimodal experiment to
  study polycrystal plasticity},
\newblock \bibinfo{journal}{Journal of Synchrotron Radiation}
  \bibinfo{volume}{30} (\bibinfo{year}{2023}) \bibinfo{pages}{379--389}.
%Type = Article
\bibitem[{King et~al.(2013)King, Reischig, Adrien, and
  Ludwig}]{king_first_2013}
\bibinfo{author}{A.~King}, \bibinfo{author}{P.~Reischig},
  \bibinfo{author}{J.~Adrien}, \bibinfo{author}{W.~Ludwig},
\newblock \bibinfo{title}{First laboratory {X}-ray diffraction contrast
  tomography for grain mapping of polycrystals},
\newblock \bibinfo{journal}{Journal of Applied Crystallography}
  \bibinfo{volume}{46} (\bibinfo{year}{2013}) \bibinfo{pages}{1734--1740}.
%Type = Article
\bibitem[{McDonald et~al.(2015)McDonald, Reischig, Holzner, Lauridsen, Withers,
  Merkle, and Feser}]{mcdonald_non-destructive_2015}
\bibinfo{author}{S.~A. McDonald}, \bibinfo{author}{P.~Reischig},
  \bibinfo{author}{C.~Holzner}, \bibinfo{author}{E.~M. Lauridsen},
  \bibinfo{author}{P.~J. Withers}, \bibinfo{author}{A.~P. Merkle},
  \bibinfo{author}{M.~Feser},
\newblock \bibinfo{title}{Non-destructive mapping of grain orientations in {3D}
  by laboratory {X}-ray microscopy},
\newblock \bibinfo{journal}{Scientific Reports} \bibinfo{volume}{5}
  (\bibinfo{year}{2015}) \bibinfo{pages}{14665}.
%Type = Article
\bibitem[{Keinan et~al.(2018)Keinan, Bale, Gueninchault, Lauridsen, and
  Shahani}]{keinan_integrated_2018}
\bibinfo{author}{R.~Keinan}, \bibinfo{author}{H.~Bale},
  \bibinfo{author}{N.~Gueninchault}, \bibinfo{author}{E.~M. Lauridsen},
  \bibinfo{author}{A.~J. Shahani},
\newblock \bibinfo{title}{Integrated imaging in three dimensions: {Providing} a
  new lens on grain boundaries, particles, and their correlations in
  polycrystalline silicon},
\newblock \bibinfo{journal}{Acta Materialia} \bibinfo{volume}{148}
  (\bibinfo{year}{2018}) \bibinfo{pages}{225--234}.
%Type = Article
\bibitem[{Zhao et~al.(2022)Zhao, Niverty, Ma, and
  Chawla}]{zhao_correlation_2022}
\bibinfo{author}{Y.~Zhao}, \bibinfo{author}{S.~Niverty},
  \bibinfo{author}{X.~Ma}, \bibinfo{author}{N.~Chawla},
\newblock \bibinfo{title}{Correlation between corrosion behavior and grain
  boundary characteristics of a 6061 {Al} alloy by lab-scale {X}-ray
  diffraction contrast tomography ({DCT})},
\newblock \bibinfo{journal}{Materials Characterization} \bibinfo{volume}{193}
  (\bibinfo{year}{2022}) \bibinfo{pages}{112325}.
%Type = Article
\bibitem[{Eguchi et~al.(2022)Eguchi, Burnett, and
  Engelberg}]{eguchi_x-ray_2022}
\bibinfo{author}{K.~Eguchi}, \bibinfo{author}{T.~L. Burnett},
  \bibinfo{author}{D.~L. Engelberg},
\newblock \bibinfo{title}{X-{Ray} {Diffraction} {Contrast} {Tomography} for
  {Probing} {Hydrogen} {Embrittlement} in {Heat}-{Treated} {Lean} {Duplex}
  {Stainless} {Steel}},
\newblock \bibinfo{journal}{Frontiers in Materials} \bibinfo{volume}{9}
  (\bibinfo{year}{2022}).
%Type = Article
\bibitem[{Miyamoto et~al.(2009)Miyamoto, Shibata, Maki, and
  Furuhara}]{MIYAMOTO20091120}
\bibinfo{author}{G.~Miyamoto}, \bibinfo{author}{A.~Shibata},
  \bibinfo{author}{T.~Maki}, \bibinfo{author}{T.~Furuhara},
\newblock \bibinfo{title}{Precise measurement of strain accommodation in
  austenite matrix surrounding martensite in ferrous alloys by electron
  backscatter diffraction analysis},
\newblock \bibinfo{journal}{Acta Materialia} \bibinfo{volume}{57}
  (\bibinfo{year}{2009}) \bibinfo{pages}{1120--1131}.
%Type = Article
\bibitem[{Zhang et~al.(2014)Zhang, Collins, Dunne, and Shollock}]{ZHANG201425}
\bibinfo{author}{T.~Zhang}, \bibinfo{author}{D.~M. Collins},
  \bibinfo{author}{F.~P. Dunne}, \bibinfo{author}{B.~A. Shollock},
\newblock \bibinfo{title}{{Crystal plasticity and high-resolution electron
  backscatter diffraction analysis of full-field polycrystal Ni superalloy
  strains and rotations under thermal loading}},
\newblock \bibinfo{journal}{Acta Materialia} \bibinfo{volume}{80}
  (\bibinfo{year}{2014}) \bibinfo{pages}{25--38}.
%Type = Article
\bibitem[{Wallis et~al.(2019)Wallis, Hansen, Britton, and
  Wilkinson}]{wallis2019}
\bibinfo{author}{D.~Wallis}, \bibinfo{author}{L.~N. Hansen},
  \bibinfo{author}{T.~B. Britton}, \bibinfo{author}{A.~J. Wilkinson},
\newblock \bibinfo{title}{{High-Angular Resolution Electron Backscatter
  Diffraction as a New Tool for Mapping Lattice Distortion in Geological
  Minerals}},
\newblock \bibinfo{journal}{Journal of Geophysical Research: Solid Earth}
  \bibinfo{volume}{124} (\bibinfo{year}{2019}) \bibinfo{pages}{6337--6358}.
%Type = Article
\bibitem[{Deal et~al.(2021)Deal, Spinelli, Chuang, Gao, and
  Broderick}]{DEAL2021111027}
\bibinfo{author}{A.~Deal}, \bibinfo{author}{I.~Spinelli},
  \bibinfo{author}{A.~Chuang}, \bibinfo{author}{Y.~Gao},
  \bibinfo{author}{T.~Broderick},
\newblock \bibinfo{title}{{Measuring residual stress in Ti-6Al-4V with HR-EBSD,
  using reference patterns from annealed material}},
\newblock \bibinfo{journal}{Materials Characterization} \bibinfo{volume}{175}
  (\bibinfo{year}{2021}) \bibinfo{pages}{111027}.
%Type = Article
\bibitem[{Koko et~al.(2023)Koko, Becker, Elmukashfi, Pugno, Wilkinson, and
  Marrow}]{KOKO2023105173}
\bibinfo{author}{A.~Koko}, \bibinfo{author}{T.~H. Becker},
  \bibinfo{author}{E.~Elmukashfi}, \bibinfo{author}{N.~M. Pugno},
  \bibinfo{author}{A.~J. Wilkinson}, \bibinfo{author}{T.~J. Marrow},
\newblock \bibinfo{title}{{HR-EBSD analysis of in situ stable crack growth at
  the micron scale}},
\newblock \bibinfo{journal}{Journal of the Mechanics and Physics of Solids}
  \bibinfo{volume}{172} (\bibinfo{year}{2023}) \bibinfo{pages}{105173}.
%Type = Article
\bibitem[{Groeber et~al.(2006)Groeber, Haley, Uchic, Dimiduk, and
  Ghosh}]{GROEBER2006259}
\bibinfo{author}{M.~Groeber}, \bibinfo{author}{B.~Haley},
  \bibinfo{author}{M.~Uchic}, \bibinfo{author}{D.~Dimiduk},
  \bibinfo{author}{S.~Ghosh},
\newblock \bibinfo{title}{{3D reconstruction and characterization of
  polycrystalline microstructures using a FIB–SEM system}},
\newblock \bibinfo{journal}{Materials Characterization} \bibinfo{volume}{57}
  (\bibinfo{year}{2006}) \bibinfo{pages}{259--273}.
%Type = Article
\bibitem[{Rohrer et~al.(2010)Rohrer, Li, Lee, Rollett, Groeber, and
  Uchic}]{Rohrer2010}
\bibinfo{author}{G.~S. Rohrer}, \bibinfo{author}{J.~Li},
  \bibinfo{author}{S.~Lee}, \bibinfo{author}{A.~D. Rollett},
  \bibinfo{author}{M.~Groeber}, \bibinfo{author}{M.~D. Uchic},
\newblock \bibinfo{title}{{Deriving grain boundary character distributions and
  relative grain boundary energies from three-dimensional EBSD data}},
\newblock \bibinfo{journal}{Materials Science and Technology}
  \bibinfo{volume}{26} (\bibinfo{year}{2010}) \bibinfo{pages}{661--669}.
%Type = Article
\bibitem[{Kalácska et~al.(2020)Kalácska, Ast, Ispánovity, Michler, and
  Maeder}]{KALACSKA2020211}
\bibinfo{author}{S.~Kalácska}, \bibinfo{author}{J.~Ast},
  \bibinfo{author}{P.~D. Ispánovity}, \bibinfo{author}{J.~Michler},
  \bibinfo{author}{X.~Maeder},
\newblock \bibinfo{title}{{3D HR-EBSD Characterization of the plastic zone
  around crack tips in tungsten single crystals at the micron scale}},
\newblock \bibinfo{journal}{Acta Materialia} \bibinfo{volume}{200}
  (\bibinfo{year}{2020}) \bibinfo{pages}{211--222}.
%Type = Article
\bibitem[{DeMott et~al.(2021)DeMott, Haghdadi, Kong, Gandomkar, Kenney,
  Collins, and Primig}]{DEMOTT2021113394}
\bibinfo{author}{R.~DeMott}, \bibinfo{author}{N.~Haghdadi},
  \bibinfo{author}{C.~Kong}, \bibinfo{author}{Z.~Gandomkar},
  \bibinfo{author}{M.~Kenney}, \bibinfo{author}{P.~Collins},
  \bibinfo{author}{S.~Primig},
\newblock \bibinfo{title}{{3D electron backscatter diffraction characterization
  of fine $\alpha$ titanium microstructures: collection, reconstruction, and
  analysis methods}},
\newblock \bibinfo{journal}{Ultramicroscopy} \bibinfo{volume}{230}
  (\bibinfo{year}{2021}) \bibinfo{pages}{113394}.
%Type = Book
\bibitem[{Schwartz et~al.(2013)Schwartz, Kumar, and Adams}]{EBSD}
\bibinfo{author}{A.~Schwartz}, \bibinfo{author}{M.~Kumar},
  \bibinfo{author}{B.~Adams}, \bibinfo{title}{{Electron Backscatter Diffraction
  in Materials Science}}, Springer series in operations research,
  \bibinfo{edition}{2} ed., \bibinfo{publisher}{Springer},
  \bibinfo{address}{New York}, \bibinfo{year}{2013}.
%Type = Article
\bibitem[{Pokharel et~al.(2020)Pokharel, Lebensohn, Pagan, Ickes, Clausen,
  Brown, Chen, Dale, and Bernier}]{Pokharel2020}
\bibinfo{author}{R.~Pokharel}, \bibinfo{author}{R.~Lebensohn},
  \bibinfo{author}{D.~Pagan}, \bibinfo{author}{T.~Ickes},
  \bibinfo{author}{B.~Clausen}, \bibinfo{author}{D.~Brown},
  \bibinfo{author}{C.-F. Chen}, \bibinfo{author}{D.~Dale},
  \bibinfo{author}{J.~Bernier},
\newblock \bibinfo{title}{{In-Situ Grain Resolved Stress Characterization
  During Damage Initiation in Cu-10\%W Alloy}},
\newblock \bibinfo{journal}{JOM} \bibinfo{volume}{72} (\bibinfo{year}{2020})
  \bibinfo{pages}{48–56}.
%Type = Article
\bibitem[{Ludwig et~al.(2009)Ludwig, Reischig, King, Herbig, Lauridsen,
  Johnson, Marrow, and Buffière}]{ludwig_three-dimensional_2009}
\bibinfo{author}{W.~Ludwig}, \bibinfo{author}{P.~Reischig},
  \bibinfo{author}{A.~King}, \bibinfo{author}{M.~Herbig},
  \bibinfo{author}{E.~M. Lauridsen}, \bibinfo{author}{G.~Johnson},
  \bibinfo{author}{T.~J. Marrow}, \bibinfo{author}{J.~Y. Buffière},
\newblock \bibinfo{title}{Three-dimensional grain mapping by x-ray diffraction
  contrast tomography and the use of {Friedel} pairs in diffraction data
  analysis},
\newblock \bibinfo{journal}{Review of Scientific Instruments}
  \bibinfo{volume}{80} (\bibinfo{year}{2009}) \bibinfo{pages}{033905}.
%Type = Article
\bibitem[{Syha et~al.(2013)Syha, Trenkle, Lödermann, Graff, Ludwig, Weygand,
  and Gumbsch}]{syha_validation_2013}
\bibinfo{author}{M.~Syha}, \bibinfo{author}{A.~Trenkle},
  \bibinfo{author}{B.~Lödermann}, \bibinfo{author}{A.~Graff},
  \bibinfo{author}{W.~Ludwig}, \bibinfo{author}{D.~Weygand},
  \bibinfo{author}{P.~Gumbsch},
\newblock \bibinfo{title}{Validation of three-dimensional diffraction contrast
  tomography reconstructions by means of electron backscatter diffraction
  characterization},
\newblock \bibinfo{journal}{Journal of Applied Crystallography}
  \bibinfo{volume}{46} (\bibinfo{year}{2013}) \bibinfo{pages}{1145--1150}.
%Type = Article
\bibitem[{Quey et~al.(2013)Quey, Suhonen, Laurencin, Cloetens, and
  Bleuet}]{quey_direct_2013}
\bibinfo{author}{R.~Quey}, \bibinfo{author}{H.~Suhonen},
  \bibinfo{author}{J.~Laurencin}, \bibinfo{author}{P.~Cloetens},
  \bibinfo{author}{P.~Bleuet},
\newblock \bibinfo{title}{Direct comparison between {X}-ray nanotomography and
  scanning electron microscopy for the microstructure characterization of a
  solid oxide fuel cell anode},
\newblock \bibinfo{journal}{Materials Characterization} \bibinfo{volume}{78}
  (\bibinfo{year}{2013}) \bibinfo{pages}{87--95}.
%Type = Article
\bibitem[{Renversade et~al.(2016)Renversade, Quey, Ludwig, Menasche, Maddali,
  Suter, and Borbély}]{renversade_comparison_2016}
\bibinfo{author}{L.~Renversade}, \bibinfo{author}{R.~Quey},
  \bibinfo{author}{W.~Ludwig}, \bibinfo{author}{D.~Menasche},
  \bibinfo{author}{S.~Maddali}, \bibinfo{author}{R.~M. Suter},
  \bibinfo{author}{A.~Borbély},
\newblock \bibinfo{title}{Comparison between diffraction contrast tomography
  and high-energy diffraction microscopy on a slightly deformed aluminium
  alloy},
\newblock \bibinfo{journal}{IUCrJ} \bibinfo{volume}{3} (\bibinfo{year}{2016})
  \bibinfo{pages}{32--42}.
%Type = Article
\bibitem[{Kang et~al.(2019)Kang, Lu, Loo, Senabulya, and
  Shahani}]{kang_polyproc_2019}
\bibinfo{author}{J.~Kang}, \bibinfo{author}{N.~Lu}, \bibinfo{author}{I.~Loo},
  \bibinfo{author}{N.~Senabulya}, \bibinfo{author}{A.~J. Shahani},
\newblock \bibinfo{title}{{PolyProc}: {A} {Modular} {Processing} {Pipeline} for
  {X}-ray {Diffraction} {Tomography}},
\newblock \bibinfo{journal}{Integrating Materials and Manufacturing Innovation}
  \bibinfo{volume}{8} (\bibinfo{year}{2019}) \bibinfo{pages}{388--399}.
%Type = Article
\bibitem[{Oddershede et~al.(2019)Oddershede, Sun, Gueninchault, Bachmann, Bale,
  Holzner, and Lauridsen}]{oddershede_non-destructive_2019}
\bibinfo{author}{J.~Oddershede}, \bibinfo{author}{J.~Sun},
  \bibinfo{author}{N.~Gueninchault}, \bibinfo{author}{F.~Bachmann},
  \bibinfo{author}{H.~Bale}, \bibinfo{author}{C.~Holzner},
  \bibinfo{author}{E.~Lauridsen},
\newblock \bibinfo{title}{Non-destructive {Characterization} of
  {Polycrystalline} {Materials} in {3D} by {Laboratory} {Diffraction}
  {Contrast} {Tomography}},
\newblock \bibinfo{journal}{Integr Mater Manuf Innov} \bibinfo{volume}{8}
  (\bibinfo{year}{2019}) \bibinfo{pages}{217--225}.
%Type = Article
\bibitem[{Oddershede et~al.(2022)Oddershede, Bachmann, Sun, and
  Lauridsen}]{oddershede_advanced_2022}
\bibinfo{author}{J.~Oddershede}, \bibinfo{author}{F.~Bachmann},
  \bibinfo{author}{J.~Sun}, \bibinfo{author}{E.~Lauridsen},
\newblock \bibinfo{title}{Advanced {Acquisition} {Strategies} for {Lab}-{Based}
  {Diffraction} {Contrast} {Tomography}},
\newblock \bibinfo{journal}{Integrating Materials and Manufacturing Innovation}
  \bibinfo{volume}{11} (\bibinfo{year}{2022}) \bibinfo{pages}{1--12}.
%Type = Article
\bibitem[{Bachmann et~al.(2019)Bachmann, Bale, Gueninchault, Holzner, and
  Lauridsen}]{bachmann_3d_2019}
\bibinfo{author}{F.~Bachmann}, \bibinfo{author}{H.~Bale},
  \bibinfo{author}{N.~Gueninchault}, \bibinfo{author}{C.~Holzner},
  \bibinfo{author}{E.~M. Lauridsen},
\newblock \bibinfo{title}{{3D} grain reconstruction from laboratory diffraction
  contrast tomography},
\newblock \bibinfo{journal}{Journal of Applied Crystallography}
  \bibinfo{volume}{52} (\bibinfo{year}{2019}) \bibinfo{pages}{643--651}.
%Type = Article
\bibitem[{Bachmann et~al.(2010)Bachmann, Hielscher, and
  Schaeben}]{bachmann_texture_2010}
\bibinfo{author}{F.~Bachmann}, \bibinfo{author}{R.~Hielscher},
  \bibinfo{author}{H.~Schaeben},
\newblock \bibinfo{title}{Texture {Analysis} with {MTEX} – {Free} and {Open}
  {Source} {Software} {Toolbox}},
\newblock \bibinfo{journal}{Solid State Phenomena} \bibinfo{volume}{160}
  (\bibinfo{year}{2010}) \bibinfo{pages}{63--68}.
%Type = Article
\bibitem[{Hielscher et~al.(2019)Hielscher, Silbermann, Schmidl, and
  Ihlemann}]{hielscher_denoising_2019}
\bibinfo{author}{R.~Hielscher}, \bibinfo{author}{C.~B. Silbermann},
  \bibinfo{author}{E.~Schmidl}, \bibinfo{author}{J.~Ihlemann},
\newblock \bibinfo{title}{Denoising of crystal orientation maps},
\newblock \bibinfo{journal}{Journal of Applied Crystallography}
  \bibinfo{volume}{52} (\bibinfo{year}{2019}) \bibinfo{pages}{984--996}.
%Type = Misc
\bibitem[{Proudhon(2021)}]{proudhon_pymicro_2021}
\bibinfo{author}{H.~Proudhon}, \bibinfo{title}{pymicro}, \bibinfo{year}{2021}.
%Type = Book
\bibitem[{Roithmayr and Hodges(2015)}]{roithmayr_dynamics_2015}
\bibinfo{author}{C.~M. Roithmayr}, \bibinfo{author}{D.~H. Hodges},
  \bibinfo{title}{Dynamics: {Theory} and {Application} of {Kane}’s {Method}},
  \bibinfo{edition}{1} ed., \bibinfo{publisher}{Cambridge University Press},
  \bibinfo{year}{2015}.
%Type = Article
\bibitem[{Miranda(2018)}]{miranda_pyswarms_2018}
\bibinfo{author}{L.~J. Miranda},
\newblock \bibinfo{title}{{PySwarms}: a research toolkit for {Particle} {Swarm}
  {Optimization} in {Python}},
\newblock \bibinfo{journal}{Journal of Open Source Software}
  \bibinfo{volume}{3} (\bibinfo{year}{2018}) \bibinfo{pages}{433}.
%Type = Inproceedings
\bibitem[{Kennedy and Eberhart(1995)}]{kennedy_particle_1995}
\bibinfo{author}{J.~Kennedy}, \bibinfo{author}{R.~Eberhart},
\newblock \bibinfo{title}{Particle swarm optimization},
\newblock in: \bibinfo{booktitle}{Proceedings of {ICNN}'95 - {International}
  {Conference} on {Neural} {Networks}}, volume~\bibinfo{volume}{4},
  \bibinfo{year}{1995}, pp. \bibinfo{pages}{1942--1948 vol.4}.
%Type = Inproceedings
\bibitem[{Shi and Eberhart(1998)}]{shi_modified_1998}
\bibinfo{author}{Y.~Shi}, \bibinfo{author}{R.~Eberhart},
\newblock \bibinfo{title}{A modified particle swarm optimizer},
\newblock in: \bibinfo{booktitle}{1998 {IEEE} {International} {Conference} on
  {Evolutionary} {Computation} {Proceedings}. {IEEE} {World} {Congress} on
  {Computational} {Intelligence} ({Cat}. {No}.{98TH8360})},
  \bibinfo{year}{1998}, pp. \bibinfo{pages}{69--73}.
%Type = Article
\bibitem[{Virtanen et~al.(2020)Virtanen, Gommers, Oliphant, Haberland, Reddy,
  Cournapeau, Burovski, Peterson, Weckesser, Bright, van~der Walt, Brett,
  Wilson, Millman, Mayorov, Nelson, Jones, Kern, Larson, Carey, Polat, Feng,
  Moore, VanderPlas, Laxalde, Perktold, Cimrman, Henriksen, Quintero, Harris,
  Archibald, Ribeiro, Pedregosa, and van Mulbregt}]{virtanen_scipy_2020}
\bibinfo{author}{P.~Virtanen}, \bibinfo{author}{R.~Gommers},
  \bibinfo{author}{T.~E. Oliphant}, \bibinfo{author}{M.~Haberland},
  \bibinfo{author}{T.~Reddy}, \bibinfo{author}{D.~Cournapeau},
  \bibinfo{author}{E.~Burovski}, \bibinfo{author}{P.~Peterson},
  \bibinfo{author}{W.~Weckesser}, \bibinfo{author}{J.~Bright},
  \bibinfo{author}{S.~J. van~der Walt}, \bibinfo{author}{M.~Brett},
  \bibinfo{author}{J.~Wilson}, \bibinfo{author}{K.~J. Millman},
  \bibinfo{author}{N.~Mayorov}, \bibinfo{author}{A.~R.~J. Nelson},
  \bibinfo{author}{E.~Jones}, \bibinfo{author}{R.~Kern},
  \bibinfo{author}{E.~Larson}, \bibinfo{author}{C.~J. Carey},
  \bibinfo{author}{I.~Polat}, \bibinfo{author}{Y.~Feng}, \bibinfo{author}{E.~W.
  Moore}, \bibinfo{author}{J.~VanderPlas}, \bibinfo{author}{D.~Laxalde},
  \bibinfo{author}{J.~Perktold}, \bibinfo{author}{R.~Cimrman},
  \bibinfo{author}{I.~Henriksen}, \bibinfo{author}{E.~A. Quintero},
  \bibinfo{author}{C.~R. Harris}, \bibinfo{author}{A.~M. Archibald},
  \bibinfo{author}{A.~H. Ribeiro}, \bibinfo{author}{F.~Pedregosa},
  \bibinfo{author}{P.~van Mulbregt},
\newblock \bibinfo{title}{{SciPy} 1.0: fundamental algorithms for scientific
  computing in {Python}},
\newblock \bibinfo{journal}{Nature Methods} \bibinfo{volume}{17}
  (\bibinfo{year}{2020}) \bibinfo{pages}{261--272}.
%Type = Book
\bibitem[{Nocedal and Wright(2006)}]{nocedal_numerical_2006}
\bibinfo{author}{J.~Nocedal}, \bibinfo{author}{S.~J. Wright},
  \bibinfo{title}{Numerical optimization}, Springer series in operations
  research, \bibinfo{edition}{2} ed., \bibinfo{publisher}{Springer},
  \bibinfo{address}{New York}, \bibinfo{year}{2006}.
%Type = Misc
\bibitem[{{ASTM International}(2021)}]{astm_international_standard_2021}
\bibinfo{author}{{ASTM International}}, \bibinfo{title}{Standard {Test}
  {Methods} for {Determining} {Average} {Grain} {Size}}, \bibinfo{year}{2021}.
%Type = Article
\bibitem[{Yousefian et~al.(2021)Yousefian, Zarei-Hanzaki, Barabi, Abedi,
  Moallemi, and Karjalainen}]{yousefian_microstructure_2021}
\bibinfo{author}{S.~Yousefian}, \bibinfo{author}{A.~Zarei-Hanzaki},
  \bibinfo{author}{A.~Barabi}, \bibinfo{author}{H.~R. Abedi},
  \bibinfo{author}{M.~Moallemi}, \bibinfo{author}{P.~Karjalainen},
\newblock \bibinfo{title}{Microstructure, texture and mechanical properties of
  a nickel-free high nitrogen duplex stainless steel processed through friction
  stir spot welding},
\newblock \bibinfo{journal}{Journal of Materials Research and Technology}
  \bibinfo{volume}{15} (\bibinfo{year}{2021}) \bibinfo{pages}{6491--6505}.
%Type = Article
\bibitem[{Zhang et~al.(2019)Zhang, Zhu, Singaravelu, Sun, Jing, and
  Chawla}]{zhang_three-dimensional_2019}
\bibinfo{author}{Q.~Zhang}, \bibinfo{author}{K.~Zhu}, \bibinfo{author}{A.~S.~S.
  Singaravelu}, \bibinfo{author}{W.~Sun}, \bibinfo{author}{T.~Jing},
  \bibinfo{author}{N.~Chawla},
\newblock \bibinfo{title}{Three-{Dimensional} ({3D}) {Microstructure}-{Based}
  {Modeling} of a {Thermally}-{Aged} {Cast} {Duplex} {Stainless} {Steel}
  {Based} on {X}-ray {Microtomography}, {Nanoindentation} and {Micropillar}
  {Compression}},
\newblock \bibinfo{journal}{Metals} \bibinfo{volume}{9} (\bibinfo{year}{2019})
  \bibinfo{pages}{688}.
%Type = Article
\bibitem[{Fu(2021)}]{fu_general_2021}
\bibinfo{author}{J.~Fu},
\newblock \bibinfo{title}{A general approach to determine texture patterns
  using pole figure},
\newblock \bibinfo{journal}{Journal of Materials Research and Technology}
  \bibinfo{volume}{14} (\bibinfo{year}{2021}) \bibinfo{pages}{1284--1291}.
%Type = Article
\bibitem[{Holzner et~al.(2016)Holzner, Lavery, Bale, Merkle, McDonald, Withers,
  Zhang, Jensen, Kimura, Lyckegaard, Reischig, and
  Lauridsen}]{holzner_diffraction_2016}
\bibinfo{author}{C.~Holzner}, \bibinfo{author}{L.~Lavery},
  \bibinfo{author}{H.~Bale}, \bibinfo{author}{A.~Merkle},
  \bibinfo{author}{S.~McDonald}, \bibinfo{author}{P.~Withers},
  \bibinfo{author}{Y.~Zhang}, \bibinfo{author}{D.~J. Jensen},
  \bibinfo{author}{M.~Kimura}, \bibinfo{author}{A.~Lyckegaard},
  \bibinfo{author}{P.~Reischig}, \bibinfo{author}{E.~Lauridsen},
\newblock \bibinfo{title}{Diffraction {Contrast} {Tomography} in the
  {Laboratory} – {Applications} and {Future} {Directions}},
\newblock \bibinfo{journal}{Microscopy Today} \bibinfo{volume}{24}
  (\bibinfo{year}{2016}) \bibinfo{pages}{34--43}.
%Type = Article
\bibitem[{Ganju et~al.(2023)Ganju, Nieto-Valeiras, LLorca, and
  Chawla}]{ganju_novel_2023}
\bibinfo{author}{E.~Ganju}, \bibinfo{author}{E.~Nieto-Valeiras},
  \bibinfo{author}{J.~LLorca}, \bibinfo{author}{N.~Chawla},
\newblock \bibinfo{title}{A novel diffraction contrast tomography ({DCT})
  acquisition strategy for capturing the {3D} crystallographic structure of
  pure titanium},
\newblock \bibinfo{journal}{Tomography of Materials and Structures}
  \bibinfo{volume}{1} (\bibinfo{year}{2023}) \bibinfo{pages}{100003}.
%Type = Article
\bibitem[{Sun et~al.(2022)Sun, Bachmann, Oddershede, and
  Lauridsen}]{sun_recent_2022}
\bibinfo{author}{J.~Sun}, \bibinfo{author}{F.~Bachmann},
  \bibinfo{author}{J.~Oddershede}, \bibinfo{author}{E.~Lauridsen},
\newblock \bibinfo{title}{Recent advances of lab-based diffraction contrast
  tomography – reconstruction speed benchmark testing and validations},
\newblock \bibinfo{journal}{IOP Conf. Ser.: Mater. Sci. Eng.}
  \bibinfo{volume}{1249} (\bibinfo{year}{2022}) \bibinfo{pages}{012045}.
%Type = Article
\bibitem[{Schindelin et~al.(2012)Schindelin, Arganda-Carreras, Frise, Kaynig,
  Longair, Pietzsch, Preibisch, Rueden, Saalfeld, Schmid, Tinevez, White,
  Hartenstein, Eliceiri, Tomancak, and Cardona}]{schindelin_fiji_2012}
\bibinfo{author}{J.~Schindelin}, \bibinfo{author}{I.~Arganda-Carreras},
  \bibinfo{author}{E.~Frise}, \bibinfo{author}{V.~Kaynig},
  \bibinfo{author}{M.~Longair}, \bibinfo{author}{T.~Pietzsch},
  \bibinfo{author}{S.~Preibisch}, \bibinfo{author}{C.~Rueden},
  \bibinfo{author}{S.~Saalfeld}, \bibinfo{author}{B.~Schmid},
  \bibinfo{author}{J.-Y. Tinevez}, \bibinfo{author}{D.~J. White},
  \bibinfo{author}{V.~Hartenstein}, \bibinfo{author}{K.~Eliceiri},
  \bibinfo{author}{P.~Tomancak}, \bibinfo{author}{A.~Cardona},
\newblock \bibinfo{title}{Fiji: an open-source platform for biological-image
  analysis},
\newblock \bibinfo{journal}{Nature Methods} \bibinfo{volume}{9}
  (\bibinfo{year}{2012}) \bibinfo{pages}{676--682}.
%Type = Article
\bibitem[{Ahrens et~al.(2005)Ahrens, Geveci, and Law}]{ahrens_paraview_2005}
\bibinfo{author}{J.~Ahrens}, \bibinfo{author}{B.~Geveci},
  \bibinfo{author}{C.~Law},
\newblock \bibinfo{title}{{ParaView}: {An} {End}-{User} {Tool} for {Large}
  {Data} {Visualization}},
\newblock \bibinfo{journal}{Visualization Handbook}  (\bibinfo{year}{2005}).
%Type = Misc
\bibitem[{{Blender Community}(2018)}]{blender_community_blender_2018}
\bibinfo{author}{{Blender Community}}, \bibinfo{title}{Blender - a {3D}
  modelling and rendering package}, \bibinfo{year}{2018}.

\end{thebibliography}
\bibliographystyle{elsarticle-num-names}

\newpage
\appendix
\section{Tomography post-processing}
\label{sec:tomo_appendix}
To present the 3D shape of the sample using the ACT data, a simple software pipeline was used.
The ACT data were reconstructed into 3D volumes by the ZEISS software and output as a sequence of TIFF images representing each vertical slice of the sample.
First, Fiji \citep{schindelin_fiji_2012} was used to import the entire image stack, crop each image to the sample region and apply a threshold based on pixel intensity to distinguish between the sample and the surrounding air.
Secondly, Paraview \cite{ahrens_paraview_2005} was used to import the thresholded image stack and generate a surface corresponding to the sample dimensions.
An STL file describing this surface was exported from Paraview, and imported into Blender \citep{blender_community_blender_2018}.
Finally, within Blender, the \emph{Decimate} modifier was used with the \emph{Planar} setting to simplify the geometry of the flat faces of the sample and reduce the number of triangles in the mesh.

\end{document}